\documentclass{jfm}
\usepackage{graphicx}
\usepackage{epstopdf, epsfig}
\usepackage{xcolor}
\usepackage{bm}
\usepackage{physics}
\usepackage{tabularx}
\usepackage{wasysym}

\shorttitle{Modelling combined size and density segregation}
\shortauthor{Y. Duan, P. B. Umbanhowar, J. M. Ottino, and R. M. Lueptow}

\title{Modelling segregation of flowing bidisperse granular mixtures varying simultaneously in size and density }

\author{Yifei Duan\aff{1},
  Paul B. Umbanhowar\aff{2},
  Julio M. Ottino\aff{1,2,3},
  \\
 \and Richard M. Lueptow\aff{1,2,3}
  \corresp{\email{r-lueptow@northwestern.edu}}
 }
\affiliation{\aff{1}Department of Chemical and Biological Engineering, Northwestern University, Evanston, Illinois 60208, USA
\aff{2}Department of Mechanical Engineering, Northwestern University, Evanston, Illinois 60208, USA
\aff{3}Northwestern Institute on Complex Systems (NICO), Northwestern University, Evanston, Illinois 60208, USA\
}

\begin{document}

\maketitle

\begin{abstract}
Flowing granular materials segregate due to differences in particle size (driven by percolation) and density (driven by buoyancy). 
Modelling the segregation of mixtures of large/heavy particles and small/light particles is challenging due to the opposing effects of the two segregation mechanisms.
Using discrete element method (DEM) simulations of combined size and density segregation we show that the segregation velocity is well described by a model that depends linearly on the local shear rate and quadratically on the species concentration. 
Concentration profiles predicted by incorporating this segregation velocity model into a continuum advection-diffusion-segregation transport model match DEM simulation results well for a wide range of particle size and density ratios.
Most surprisingly, the DEM simulations and the segregation velocity model both show that the segregation direction for a range of size and density ratios depends on the local species concentration. 
This leads to a methodology to determine the combination of particle size ratio, density ratio, and particle concentration for which a bidisperse mixture will not segregate.
\end{abstract}


\section{Introduction}
\label{section1}

Granular segregation has been widely studied due to its importance in many areas ranging from geophysics to industrial processes \citep{ottino2008mixing,gray2010large,gray2011multi,gray2018particle,umbanhowar2019modeling}. 
Among the particle properties that drive segregation, size and density are usually the dominant factors \citep{savage1988particle,khakhar1997radial}.
In dense granular flows of size-disperse particles having the same density (S-systems), large particles tend to rise as small particles fall through voids, a segregation mechanism known as percolation \citep{williams1968mixing,drahun1983mechanisms,savage1988particle,ottino2000mixing}. 
For density-disperse mixtures of equal diameter particles (D-systems), segregation is driven by a buoyant force mechanism in which heavy particles sink and light particles rise \citep{ristow1994particle,khakhar1997radial,khakhar1999mixing,pereira2011insights}.
Various continuum models have been proposed for segregation in S- or D-system \citep{gray2018particle,umbanhowar2019modeling}, but few studies focus on bidisperse mixtures where the constituent species vary in both size and density (i.e., SD-systems). 
Though size and density differences can reinforce each other, e.g., in mixtures of large light (LL) particles and small heavy (SH) particles, a greater challenge is to predict segregation when the two segregation mechanisms oppose each other, e.g., in mixtures of large heavy (LH) particles and small light (SL) particles.
The goal here is to model particle segregation in a two-species mixture of particles differing simultaneously in both particle size and density. 

Previous studies used experiments and particle-based simulations to determine the crossover condition between percolation and buoyancy for SD-systems.
For vibrated granular materials (the Brazil nut problem), segregation occurs as a consequence of periodic dilation and compaction of the particle bed. The tendency of particles to sink or rise in vibrated systems can be characterized by the ratios of large to small particle diameter, $R_d=d_l/d_s$, and density, $R_\rho=\rho_l/\rho_s$ \citep{hong2001reverse,breu2003reversing,jenkins2002segregation,ciamarra2006granular}. 
These studies show that percolation dominates (i.e., large particles rise) for $R_d>R_\rho$.
On the other hand, in flowing granular systems segregation is related to the local shear and resulting dilation in the relatively thin gravitationally-driven flowing layers that are ubiquitous in industrial settings such as heaps \citep{fan2017segregation}, chutes \citep{savage1988particle,pouliquen1999scaling}, and tumblers \citep{hill1999segregation,liu2013effect}, as well as in geophysical flows such as landslides \citep{johnson2012grain}.
For example, free-surface flow experiments with an equal-volume mixture of large steel and small glass particles ($R_\rho=3$) in a rotating tumbler show that segregation diminishes for $R_d>2$ \citep{jain2005combined,jain2005regimes}.
In addition to size and density ratios, the crossover condition in rotating tumbler experiments also depends on particle species concentration \citep{alonso1991optimum}, such that LH particles sink at low concentrations but float at high concentrations.

Segregation experiments demonstrate the subtle interactions between particle size and density, and DEM simulations reproduce these results for laboratory-scale geometries \citep{fan2013kinematics,fan2014modelling, combarros2014segregation,xiao2016modelling,garcia2016segregation}. However, a first-principles-based predictive theory for combined size and density segregation is lacking. 
Efforts have been made to extend continuum mixture theory \citep{atkin1976continuum} beyond segregation in S- and D-systems \citep{bridgwater1985particle,dolgunin1995segregation,gray2005theory,fan2011theory} to model segregation in SD-systems \citep{marks2012grainsize,tunuguntla2014mixture,gray2015particle}. The fundamental mechanisms on which these models are formulated (i.e., partial stresses, interspecies drag) are not fully understood, and studies show that some assumptions in the theory do not match results obtained from DEM simulations \citep{weinhart2013discrete,tunuguntla2017comparing,duan2020segregation}.
A more fundamental approach is based on inter-particle interactions (i.e., Kinetic Theory of Granular Flows) \citep{arnarson2004binary,larcher2013segregation,larcher2015evolution}. However, this approach is limited to mixtures that differ little in particle size or mass and it tends to underestimate the segregation rate \citep{larcher2015evolution}.

As an alternative, we consider a transport equation approach combined with a mixture-specific segregation velocity model that has previously been used to predict either size or density segregation alone \citep{fan2014modelling,xiao2016modelling}.
This approach has been successfully applied to various flow geometries as well as multi- and polydisperse particle distributions \citep{umbanhowar2019modeling}, but has not yet been applied to SD-systems where the particles species differ in both size and density. 
In this model, the local segregation velocity of species $i$ normal to the free surface is defined as $w_{seg,i}=w_i-w$, where $w$ is the bulk surface-normal velocity, $w_i$ is surface-normal velocity of species $i$ and concentration-gradient-driven effects are ignored.
For bidisperse mixtures of non-cohesive mm-sized particles varying in a \textit{single} property (either size or density), previous studies \citep{fan2014modelling,schlick2015modeling,xiao2016modelling} show that $w_{seg,i}$ (previously referred as $w_{p,i} $ where the $p$ subscript is a remnant of the initial application to the percolation velocity in size-disperse systems) can be modelled with reasonable fidelity using
\begin{equation}
w_{seg,i}=A d_s\dot\gamma (1-c_i),
\label{wps}
\end{equation}
where $d_s$ is the small particle diameter ($d_s=d_l$ for D-systems), $\dot\gamma$ is the local shear rate, $1-c_i$ is the local concentration of the other species comprising the mixture, and the segregation coefficient, $A$, is a function of particle size or density ratio for the two species.
Equation~(\ref{wps}) provides an accurate description of the segregation velocity in most situations, but can fall short under certain conditions. 
First, recent studies indicate that segregation flux, $\Phi_{seg,i}=w_{seg,i}c_i$, has an underlying asymmetry \citep{van2015underlying,jones2018asymmetric} that depends on local particle concentration [i.e., small (or heavy) particles among mostly large (or light) particles segregate faster than the other way around], whereas equation~(\ref{wps}) predicts a segregation flux dependence on concentration that is symmetric with respect to $c_i=0.5$. 
Second, equation~(\ref{wps}) does not consider size and density ratios simultaneously. That is, $A$ has been expressed as a function of either the size ratio alone \citep{schlick2015modeling} [or volume ratio for non-spherical particles \citep{jones2020remarkable}] in S-systems or the density ratio alone \citep{xiao2016modelling} in D-systems, but not both. 
As granular materials of practical interest can vary in both size and density, a more general segregation velocity model is needed. 

To address the observed asymmetry of the segregation velocity at equal concentrations for size segregation in chute flows, \citet{gajjar2014asymmetric} proposed a flux model equivalent to a two-parameter quadratic form for the segregation velocity,
\begin{equation}
w_{seg,i}=A_\kappa(1-c_i)(1-\kappa c_i),
\label{kappa}
\end{equation}
where $A_\kappa$ is a magnitude coefficient and $\kappa$ is an asymmetry coefficient. 
\citet{jones2018asymmetric} showed that this model characterizes both size- and density-bidisperse mixtures over a wide range of size or density ratios, and expressed equation~(\ref{kappa}) in a form consistent with the linear segregation velocity model of equation~(\ref{wps}) as
\begin{equation}
w_{seg,i}= [A_{i}+B_{i}(1-c_i)]d_s\dot\gamma(1-c_i),
\label{quadkappa}
\end{equation}
where $A_i$ and $B_i$ can be determined from bidisperse heap flow simulations for a wide range of size or density ratios.
Comparison of equations~(\ref{wps}) and (\ref{quadkappa}) is enlightening. 
Both forms depend on the product of the small particle diameter, the local shear rate, and the concentration of the other species, i.e., $d_s\dot\gamma(1-c_i)$. 
Furthermore, the quadratic model has a two-parameter, concentration dependent term $A_{i}+B_{i}(1-c_i)$ instead of the concentration independent coefficient $A$ in the linear model. 

In order to model combined size and density segregation, it is necessary to find an expression for the segregation velocity in a bidisperse mixture of particles accounting for both particle size and density. 
To do this, we perform DEM simulations of particle mixtures having a wide range of particle properties, varying in both size and density, in quasi-2D bounded heap flow. 
The advantage of this flow geometry is that it is simple to implement, is steady in a frame of reference rising with the heap surface, and generates a wide range of segregation velocities, shear rates, and concentrations in a single simulation. 
Because of our previous success with equation~(\ref{quadkappa}), we focus on this model for the combined size and density segregation, knowing that this equation accurately reflects the segregation velocity in the limits of pure size-driven or pure density-driven segregation over a range of concentrations.

DEM simulation of quasi-2D single-sided bounded heap flow for size and density bidisperse mixtures is described in \S\ref{section2}.  
Based on simulation results, a segregation velocity model analogous to equation~(\ref{quadkappa}) is extended to SD-systems in \S\ref{section3}.
Using this model, it is possible to predict the combination of size ratio, density ratio, and concentration that minimizes segregation, an important result in practical engineering systems to assure that particle mixtures remain mixed.
In \S\ref{section4}, the continuum advection-diffusion-segregation transport equation is solved using the combined size and density segregation model to demonstrate that the model predictions match the simulation results for a range of feed rates, feed concentrations, and heap geometries. Conclusions are given in \S\ref{section5}.

\section{Simulation}
\label{section2}

\begin{figure}
    \centerline{\includegraphics[width=4 in]{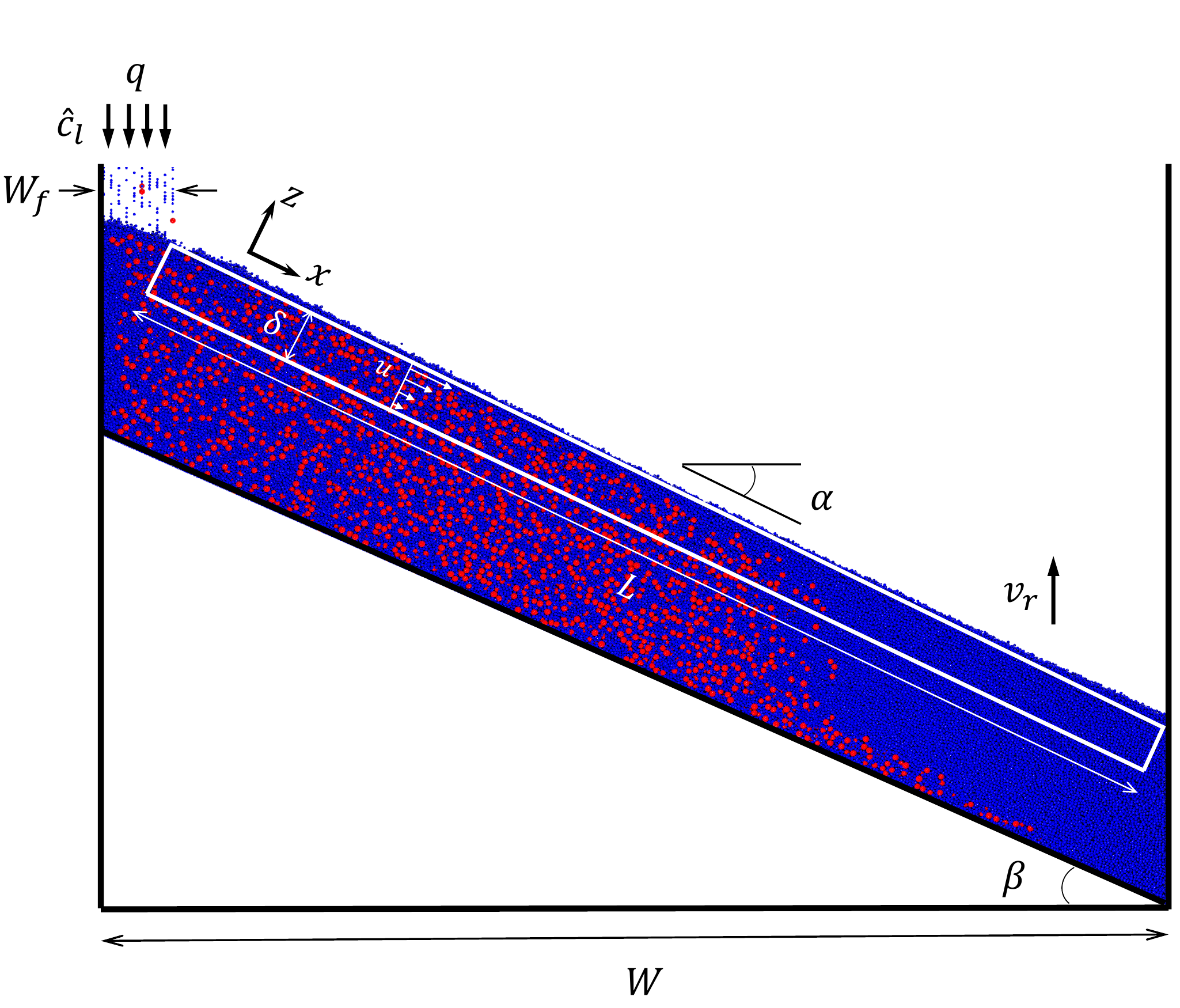}}
    \caption{Quasi-2D bounded heap simulation setup and segregation example. For these conditions, large, heavy particles (\textcolor{red}{red}, $d_l=3\,$mm, $\rho_l=4\,$g/cm$^3$) sink while small, light particles (\textcolor{blue}{blue}, $d_s=1.5\,$mm, $\rho_s=1\,$g/cm$^3$) rise. Segregation occurs in the flowing layer, which is outlined schematically by the white rectangle (the thickness is exaggerated by a factor of about two to make it more visible). $R_d=2$, $R_\rho=4$, $W_f=3.3\,$cm, $W=50\,$cm, $L=52\,$cm, $q=20\,$cm$^2/$s, $\delta=1.5\,$cm, and large particle feed concentration $\hat c_l=0.2$.  
     }
    \label{scheme}
\end{figure}

We numerically simulate combined size and density segregation of bidisperse mixtures in a single-sided quasi-2D bounded heap where particles flow in a thin surface layer down a slope much like the flow that occurs when filling a silo. 
An advantage of heap flows over other flow configurations (e.g., plane shear flows and chute flows) is that the local shear rate and the particle species concentration vary throughout the length and depth of a steady flowing layer but remain constant at a particular location in the flow (when analyzed from a reference frame that rises with the heap surface), such that time-averaged segregation data for a wide range of flow conditions can be obtained from just one simulation. 
Segregation model parameters obtained from the steady single-sided quasi-2D bounded heap geometry are universal in that they can be applied to unsteady flows and other flow geometries \citep{schlick2015granular,xiao2019continuum,deng2019modeling,isner2020axisymmetric,deng2020modeling}.
In this study, we conduct more than 350 simulations with different combinations of particle size and density ratios in developing the SD-disperse segregation model.

Our in-house DEM code \citep{isner2020axisymmetric,isnergranular} runs on CUDA-enabled GPUs and has been previously validated by heap flow experiments with mm-sized particles \citep{xiao2016modelling,isnergranular}. 
The bounded heap example shown in figure~\ref{scheme} is confined by two parallel plates in the spanwise (normal to the $xz$-plane) direction with a gap thickness of $T=15\,$mm.
The heap width between the bounding endwalls is $W=0.5\,$m.
To reduce the number of simulated particles and save computation time, the bottom wall is inclined at an angle $\beta=28^\circ$, roughly matching the dynamic repose angle $\alpha$ in steady state.
To create a rough bottom boundary, particles that contact the bottom wall are immobilized. 
After the particle bed exceeds 10-15 particle diameters in depth, the velocity and concentration profiles in the flowing layer become steady, indicating that effects of the bottom boundary can be neglected. 
A linear spring-dashpot interaction model is used to calculate inter-particle collisional forces at the point of contact \citep{cundall1979discrete}.
For a pair of colliding spheres (indicated by subscripts $j$, $k$) the normal contact force is $\bm F^n_{jk}=[k_n \epsilon -2\gamma_n m_{eff}( \bm V_{jk} \cdot \hat{\bm r}_{jk} )] \hat{\bm r}_{jk}$, where $\hat{\bm r}_{jk}$ is the normal vector, $m_{eff}=m_km_l/(m_k+m_l)$ is the effective mass, $\epsilon$ and $\bm V_{jk}$ are the overlap and relative velocity between the two spheres. The normal stiffness $k_n=[(\pi/t_c)^2+\gamma_n^2]m_{eff}$ and damping $\gamma_n=-\ln(e)/t_c$ are determined from the restitution coefficient $e$ and binary collision time $t_c$. 
 The tangential force $\bm F^t_{jk}=-\min(  |k_s\beta|,|\mu \bm F^n_{jk}|  ) \text{sgn}(\beta) \hat{\bm s}$ is estimated by a linear spring model with Coulomb friction \citep{shafer1996force}.
Here $\bm {\hat s}$ is the unit vector in the tangential direction.
The tangential stiffness is $k_s=\frac{2}{7}k_n$, and the tangential displacement is $\beta(t)=\int_{t_s}^{t}\bm V_{jk}^s dt$, where $t_s$ is the initial contact time and $\bm V_{jk}^s$ is the relative tangential velocity.
For all simulations, particle-particle and particle-wall contacts use a friction coefficient of $\mu=0.4$,
a binary collision time of $t_c=0.5\,$ms, and a restitution coefficient of $e=0.2$. 
Previous results indicate that segregation results like those considered here are largely independent of the friction and restitution coefficients \citep{jing2017micromechanical,duan2020segregation,jing2020rising}.
To fully resolve particle collisions, the DEM simulation time step is $t_c/50$ \citep{silbert2001granular,duan2017incorporation,duan2019new}.

\begin{table}
  \begin{center}
\def~{\hphantom{0}}
  \begin{tabular}{ll}
      $W$ & ~~~~~0.4, 0.5$\,$m \\
          $\mu$ & ~~~~~$0.4$ \\
   $e$ & ~~~~~$0.2$ \\
     $t_c$ & ~~~~~0.5\,ms  \\
               $\hat c_l$ & ~~~~~0.2\,--\,0.8   \\
     $q$ & ~~~~~15\,--\,20\,cm$^2$/s \\
          $R_d=d_l/d_s$    & ~~~~~1.5\,--\,2.5 ($d_l=3\,$mm)\\
     $ R_\rho=\rho_l/\rho_s$  & ~~~~~0.25\,--\,4  ($\rho_s=1\,$g/cm$^3$)\\
  \end{tabular}
   \caption{Simulation parameters.}  
   \label{table1}
  \end{center}
\end{table}

Flow conditions and particle properties differentiated by subscript $i$ ($i=l$ for large particles and $i=s$ for small particles regardless of their densities) are listed in Table~\ref{table1}. 
For a given species, the particle diameter $d_i$ is uniformly distributed with a variance of $\pm 0.1d_i$ to reduce ordering, except where noted.
A well-mixed bidisperse stream of particles with a feed concentration of large particles, $\hat c_l$, (and corresponding small particle feed concentration $\hat c_s=1-\hat c_l$) is continuously fed into the system from a relatively low height of  4$\,$cm above the rising free surface to reduce bouncing in a $3.3\,$cm (11$d_l$) long feed-zone ($W_f$).
Based on volume conservation, the free surface rises at a vertical rise velocity of $v_r=Q/WT$, where $Q$ is the volumetric feed rate. 
The flowing layer length is $L=W/$cos$(\alpha)$, and an effective 2D feed rate is defined as $q=Q/T$. 

In a reference frame rising with the flowing layer, the origin of the coordinate system is located on the free surface at the front wall at the downstream (right) edge of the vertical feed region with $x$, $y$, and $z$ oriented in the streamwise, spanwise, and normal directions, respectively.
To characterize the flow, the velocity field, $\pmb u=u \hat x +v \hat y +w \hat z$ (noting that no subscript is used for variables representing the mixture) and species concentration, $c_i$, are calculated from spatial and temporal averages of simulation data in the flowing layer.
To compute the spatial average, we use a volume-weighted binning method \citep{fan2013kinematics} with right cuboid bins oriented with two faces parallel to the free surface, two faces perpendicular to the free surface, and two faces parallel to the sidewalls. Each bin has a streamwise length of $1\,$cm (3.33$d_l$), and a height (normal to the free surface) of $1\,$mm (0.33$d_l$). 
Since particles can overlap multiple bins, the partial volumes of particles are applied to the appropriate bin for averaging purposes.
As such, the species concentration is defined as
\begin{equation}
c_i=\frac{\sum _{k\in i}^{N_i} V_{k} }{\sum _{k=1}^{N} V_k },
\label{ci}
\end{equation}
where $N_i$ and $N$ are the number of particles of species $i$ and the total number of particles in the bin, respectively, and $V_{k}$ is the volume of particle $k$ in the bin. 
The mean velocity of species $i$ is calculated as the sum of volume-weighted velocities,
\begin{equation}
\pmb u_{i}=\frac{\sum _{k\in i}^{N_i} \pmb u_{k} V_{k} }{\sum _{k\in i}^{N_i} V_k},
\label{ui}
\end{equation}
where $\pmb u_{k}$ is the vector velocity of particle $k$. 
The bulk flow velocity, $\pmb u$, is determined as $\pmb u=\sum \pmb u_ic_i$. 
To perform the temporal average, the concentration and velocity values of each bin are averaged over 5\,s at intervals of $0.01\,$s after flow reaches steady state. 
Note that an alternative expression for the bulk velocity, $\pmb u=\sum \pmb u_i \rho_i c_i/\sum \rho_i c_i$, can be calculated based on the sum of mass-weighted velocities instead of volume-weighted velocities in equation~(\ref{ui}).
In previous research utilizing the framework of mixture theory \citep{gray2015particle}, the segregation velocity was derived from mass and momentum conservation, which requires an evolving bulk density along with mass-weighted mean velocities for density disperse mixtures.
Here we assume that volume is approximately conserved, which is equivalent to assuming a nearly constant volume fraction. This allows us to use $\pmb u=\sum \pmb u_i c_i$ instead of mass-weighted velocities. And because heap flow kinematics are nearly independent of particle size and density and species concentration as shown in \S\ref{section4}, it is possible to use a volume-based transport equation to model the segregation.

In quasi-2D bounded heap flow, segregation mainly occurs in the $z$-direction (normal to the free surface), as noted in previous studies \citep{fan2014modelling,schlick2015granular,xiao2016modelling,deng2018continuum}. 
Furthermore, an advection-diffusion transport equation has been successfully used to model the segregation \citep{bridgwater1985particle,dolgunin1995segregation,gray2018particle,umbanhowar2019modeling}. 
Within this continuum framework, the concentration of species $i$ can be expressed as 
\begin{equation}
\frac{\partial c_i}{\partial t} + {\div (\pmb u^*_i c_i)}={\div (D\nabla c_i)}.
\label{transport}
\end{equation}
Here, the local collisional diffusion coefficient $D$ is a scalar, although in general it is a tensor. This approximation is accurate for flows with a single dominant shear direction \citep{umbanhowar2019modeling}. $\pmb u^*_i$ represents the diffusionless mean velocity of species $i$, which differs from the overall mean velocity, $\pmb u_i$, determined from simulation according to equation~(\ref{ui}). 
Since there is no net motion of species in the spanwise ($y$) direction (i.e. zero spanwise velocity $v_i=0$), the other velocity components of species $i$ are written most generally as $ u^*_i= u + u_{seg,i}$ and $w^*_i=w+w_{seg,i}$ where $u_{seg,i}$ and $w_{seg,i}$ are the components of the gravity-driven segregation velocity of species $i$ relative to the mean flow velocity. However, for the quasi-2D bounded heap and most other free surface flows, $u_{seg,i} \ll u$ so that $u^*_i$ can be accurately approximated by $u$ \citep{deng2018continuum}. With these assumptions, equation~(\ref{transport}) can be written as
\begin{equation}
\frac{\partial c_i}{\partial t} + \frac{\partial{u c_i}}{\partial x}+\frac{\partial  (w+w_{seg,i})c_i}{\partial z}=\frac{\partial}{\partial x} \Big( D\frac{\partial c_i}{\partial x} \Big)+\frac{\partial}{\partial z} \Big( D\frac{\partial c_i}{\partial z} \Big),
\label{transport1}
\end{equation}
or, rearranging, as
\begin{equation}
\frac{\partial c_i}{\partial t} + \frac{\partial{u c_i}}{\partial x}+\frac{\partial  \big[wc_i+w_{seg,i}c_i-D\frac{\partial c_i}{\partial z} \big] }{\partial z}=\frac{\partial}{\partial x} \Big( D\frac{\partial c_i}{\partial x} \Big).
\label{transport2}
\end{equation}
When the normal component of flux for species $i$ is measured from DEM simulation, it is the entire quantity within the brackets of equation (\ref{transport2}) that is measured.
In other words, the measured normal flux $\Phi_{i}=w_{i}c_i$ is driven by three distinct mechanisms: advection ($\Phi_{adv,i}=wc_i$), segregation ($\Phi_{seg,i}=w_{seg,i}c_i$), and diffusion ($\Phi_{D,i}=-D\partial c_i/\partial z$), and can be written as
\begin{equation}
w_{i}c_i=wc_i+w_{seg,i}c_i-{D}\frac{\partial c_i}{\partial z}.
\label{wpo}
\end{equation}
Previous studies indicate that $\Phi_{D,i}$ is typically small compared to the segregation flux, $\Phi_{seg,i}$ (i.e., $\Phi_{D,i}<0.1\Phi_{seg,i}$) for size (or density) only segregation \citep{jones2018asymmetric}.
As such, the segregation velocity is expressed simply as $w_{seg,i}\approx w_{i}-w$ \citep{fan2014modelling,schlick2015modeling,jones2018asymmetric}.
However, for the combined size and density segregation considered here, the opposing effects of size and density differences can result in very weak segregation. As a result, the concurrent concentration gradient driven diffusion flux, $\Phi_{D,i}$, can be similar in magnitude to the segregation flux, $\Phi_{seg,i}$. Thus, it is necessary to include all three terms when calculating the segregation velocity.
Using equations~(\ref{ci}) and (\ref{ui}), $w_i$, $w$, and $c_i$ can be readily calculated from simulations, so the segregation velocity can be determined as 
\begin{equation}
w_{seg,i}=w_{i}-w+\frac{1}{c_i} D\frac{\partial c_i}{\partial z} .
\label{wpi}
\end{equation}
Like all the other variables on the r.h.s.~of equation~(\ref{wpi}), the diffusion coefficient $D$ is determined directly from simulation as the mean-square displacement in the normal direction of every particle in a bin over a period $\Delta t$ \citep{utter2004self,wandersman2012particle,fan2015shear},
\begin{equation}
MSD_z(\Delta t)=\frac{1}{N}\sum_{k=1}^{N}[z_k(t+\Delta t)-z_k(t)-L(\Delta t)]^2.
\label{MSD}
\end{equation}
Here $z_k(t+\Delta t)-z_k(t)$ is the displacement of particle $k$ in the bin in a time interval $\Delta t$. $L(\Delta t)$ is the mean cumulative displacement of particles in the bin due to the bulk flow in the $z$-direction.
The $MSD_z$ values of each bin for $\Delta t$ are averaged over 200 distinct times $t$ at intervals of 0.25\,s, consistent with the 5\,s sampling window for calculating the concentration and velocity fields.
Similar to previous results \citep{fan2015shear}, $MSD_z$ data are linear in $\Delta t$ for $0.05\,\text{s}<\Delta t<0.3\,$s, indicating diffusive behaviour. 
The diffusion coefficient $D$ is then estimated as one-half the slope of a linear fit of $MSD_z$ versus $\Delta t$ \citep{utter2004self,fan2015shear}.
Further details are provided in the Supplementary Material.

\section{Segregation velocity}
\label{section3}

\begin{figure}
    \centerline{\includegraphics[width=5 in]{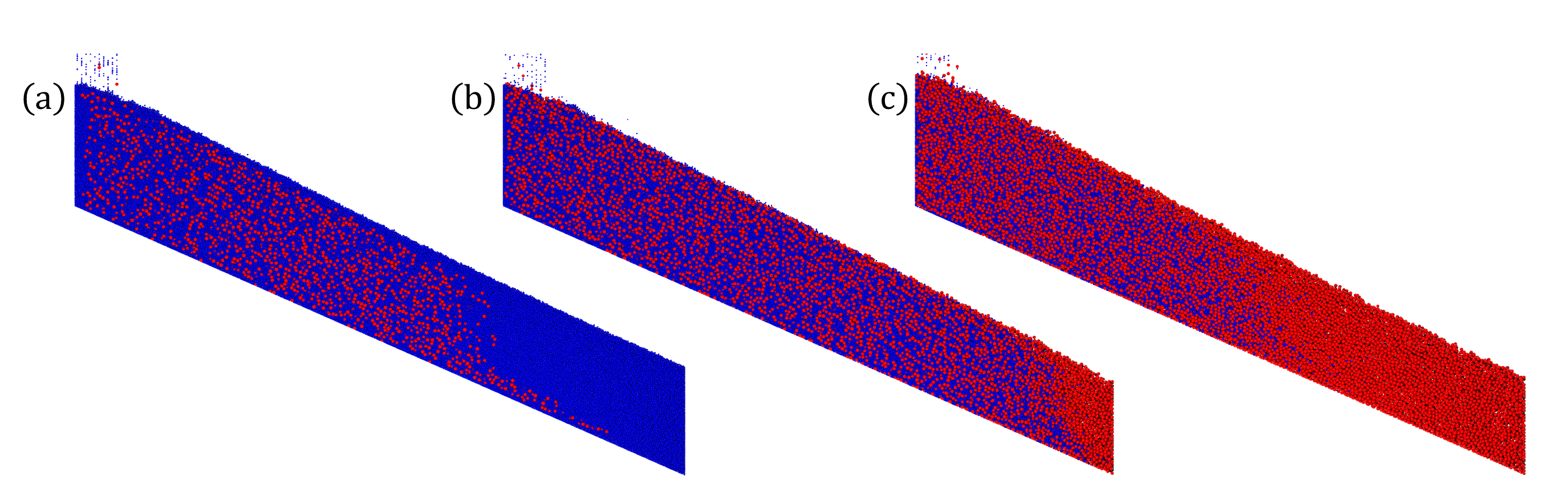}}
    \caption{Heap flow segregation for large particle feed concentration $\hat c_l$ of (a) 0.2, (b) 0.5, and (c) 0.8. Large, heavy (LH) particles (\textcolor{red}{red}, $d_l=3\,$mm, $\rho_l=4\,$g/cm$^3$) sink while small, light (SL) particles (\textcolor{blue}{blue}, $d_s=1.5\,$mm, $\rho_s=1\,$g/cm$^3$) rise for $\hat c_l=$0.2, as buoyancy overcomes percolation. In contrast, for $\hat c_l=$0.5 and 0.8 segregation reverses as percolation dominates over buoyancy. $R_d=2$, $R_\rho=4$, $W=50\,$cm, $q=20\,$cm$^2/$s.  
     }
    \label{comparison}
\end{figure}

To illustrate the interplay between size and density, consider a bidisperse mixture of particles with $R_d=2$ and $R_\rho=4$. Since in this case the large particles are also heavier particles, size and density segregation oppose one another. It is also a case in which the rising and sinking behaviour of each species has been shown to depend on the relative concentration of the two species based on experiments in a rotating tumbler \citep{alonso1991optimum}.
In their experiments, LH particles segregate to the core of the tumbler bed at low global (mixture) concentrations and segregate to the periphery of the bed at high concentrations. 

Figure~\ref{comparison} shows the analogous situation in DEM simulations of the same bidisperse particle mixture in bounded heap flow. Varying the feed concentration of large particles $\hat c_l$ significantly alters the composition of the mixture deposited on the heap just as it does for the rotating tumbler experiments.
Specifically, LH particles deposit in the upstream portion of the heap for low feed concentrations (analogous to segregating to the core of the tumbler) and deposit in the downstream portion of the heap at high feed concentrations (analogous to segregating to the tumbler periphery). 
As is shown below in more detail, this reversal in behaviour occurs because the local segregation flux of the two species depends on their local concentrations. 
This dependence of the segregation flux on concentration is different from that in either S- or D-systems. 
In S-systems, the greatest segregation flux occurs for large particle concentrations of about 0.6 for this diameter ratio ($R_d=2$).
Likewise, for D-systems, the greatest segregation flux occurs for heavy particle concentrations of about 0.4 for this density ratio ($R_\rho=4$) \citep{jones2018asymmetric}.
For the cases shown in figure~\ref{comparison} the direction of the segregation flux depends on concentration.
LH particles segregate downward at low concentrations and segregate upward at high concentrations.
Hence, in figure~\ref{comparison}(a) the low concentration LH particles segregate downward and deposit along with fewer SL particles on the upstream portion of heap until LH particles are depleted, leaving only SL particles. 
In figure~\ref{comparison}(c), the high concentration LH particles segregate upward so that the SL particles deposit along with fewer LH particles on the upstream portion of the heap until the SL particles are depleted. Figure~\ref{comparison}(b) shows an intermediate case where percolation is only slightly stronger than buoyancy.

To quantify the segregation for the cases shown in figure~\ref{comparison}, it is necessary to use equation~(\ref{wpi}) to find the segregation velocity, which requires knowledge of $c_i$, $w_i$, $w$, and $D$ found as functions of position using equations~(\ref{ci}), (\ref{ui}), and (\ref{MSD}).
Segregation occurs in the surface layer having length $L=W/$cos$(\alpha)$ and local thickness $\delta(x)$, which is defined here as the depth at which the streamwise velocity is 1/10th its surface value, i.e. $u(x,-\delta)=0.1u(x,0)$. The location of the flow surface at each streamwise position is estimated based on a cutoff value of solids fraction $\phi_c=0.35$ \citep{fan2013kinematics} to exclude bouncing particles near the free surface.

\begin{figure}
    \centerline{\includegraphics[width=5.1 in]{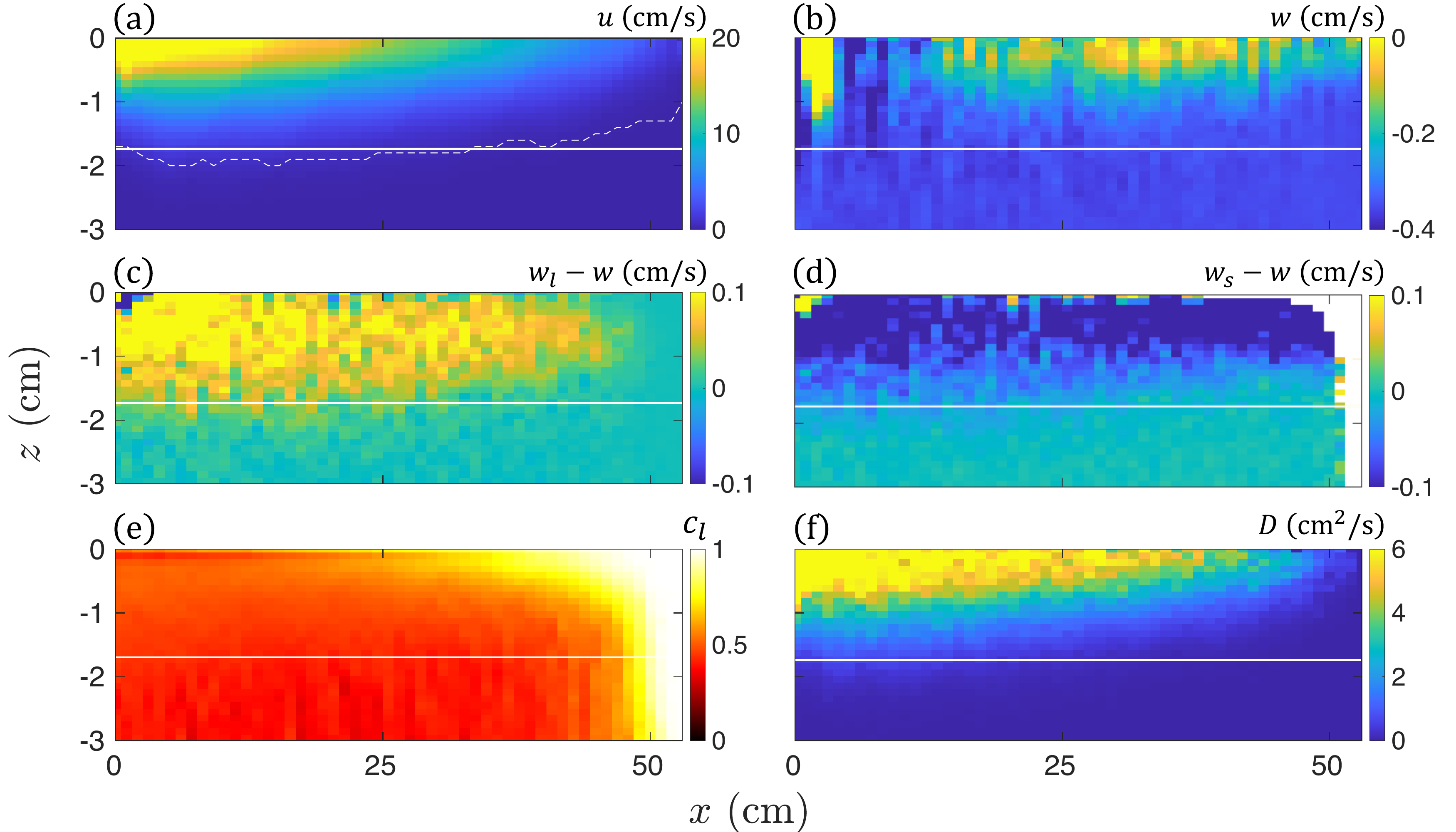}}
    \caption{ Spatial distributions of various quantities for an example simulation with $R_d=2$, $R_\rho=4$, $\hat c_l=0.5$ [i.e., data from figure~\ref{comparison}(b) rotated by repose angle $\alpha$]. Average fields of (a) streamwise velocity, (b) normal velocity (note that particles deposited on the bed have a velocity of $-v_r\cos{\alpha}=-0.35\,$cm/s due to the rising reference frame), relative normal velocities of (c) large and (d) small particles, (e) large particle concentration, and (f) diffusion coefficient. Dashed curve in (a) represents $\delta(x)$ using the criterion $u(x,-\delta)=0.1u(x,0)$. Region above solid white line in (a-f) corresponds to the constant depth flowing layer defined by the average flowing layer depth $\delta=\langle\delta(x)\rangle$.}
    \label{contour}
\end{figure}

The streamwise velocity, surface-normal velocity, species-specific velocity relative to the bulk, large particle concentration, and collisional diffusion coefficient, which are needed to calculate and model the segregation, are shown in figure~\ref{contour}.
The rectangular region above the white line corresponds to the flowing layer shown schematically in figure~\ref{scheme}. 
Figure~\ref{contour}(a) shows the local streamwise velocity, which is greatest at the surface and decreases moving downstream and deeper in the flow. 
The local flowing layer depth, $\delta(x)$, shown by the dashed curve, remains almost constant for most of the length of flowing layer, except near the downstream bounding endwall where it decreases slightly. 
Although a varying flowing layer thickness can be implemented in the continuum segregation model \citep{isner2020axisymmetric},  
a constant flowing layer depth $\delta=\langle\delta(x)\rangle$ is assumed later in this paper, as the spatial average is easier to implement and provides sufficient accuracy to successfully apply the theory for the quasi-2D heap flows considered here \citep{fan2014modelling,xiao2016modelling}.
Figure~\ref{contour}(b) shows that the normal velocity at the bottom of the flowing layer in the rising reference frame is opposite and approximately equal to the rise velocity of the surface of the heap, $v_r\cos(\alpha)=0.35\,$cm/s, except for the upstream portion near the feed-zone, where the kinematics are affected by falling particles from the vertical feed. 
Figure~\ref{contour}(c) shows that large particles rise to the surface as their normal relative velocity is positive over most of the flowing layer. 
The normal relative velocity of the large particles is zero for the downstream portion of the flowing layer corresponding to the region devoid of small particles and hence no relative velocity between the large particles and the bulk flow.
On the other hand, the relative velocity is negative for small particles shown in figure~\ref{contour}(d) except very near the downstream endwall where no small particles are present.
In other words, in the limit of $c_i=0$, the definition of segregation velocity loses its physical meaning. To account for this deficiency in the mathematical description, only data for $0.01<c_i<0.99$ are considered in the later analysis.   
The concentration of large particles, shown in figure~\ref{contour}(e), is close to 1 at the surface of the flowing layer and downstream where the entire flowing layer thickness is made up of large particles. 
As shown in figure~\ref{contour}(f), the diffusion coefficient is largest near the surface in the upstream portion of the flowing layer and decreases moving downstream and deeper into the flowing layer. 

\begin{figure}
    \centerline{\includegraphics[width=5.2 in]{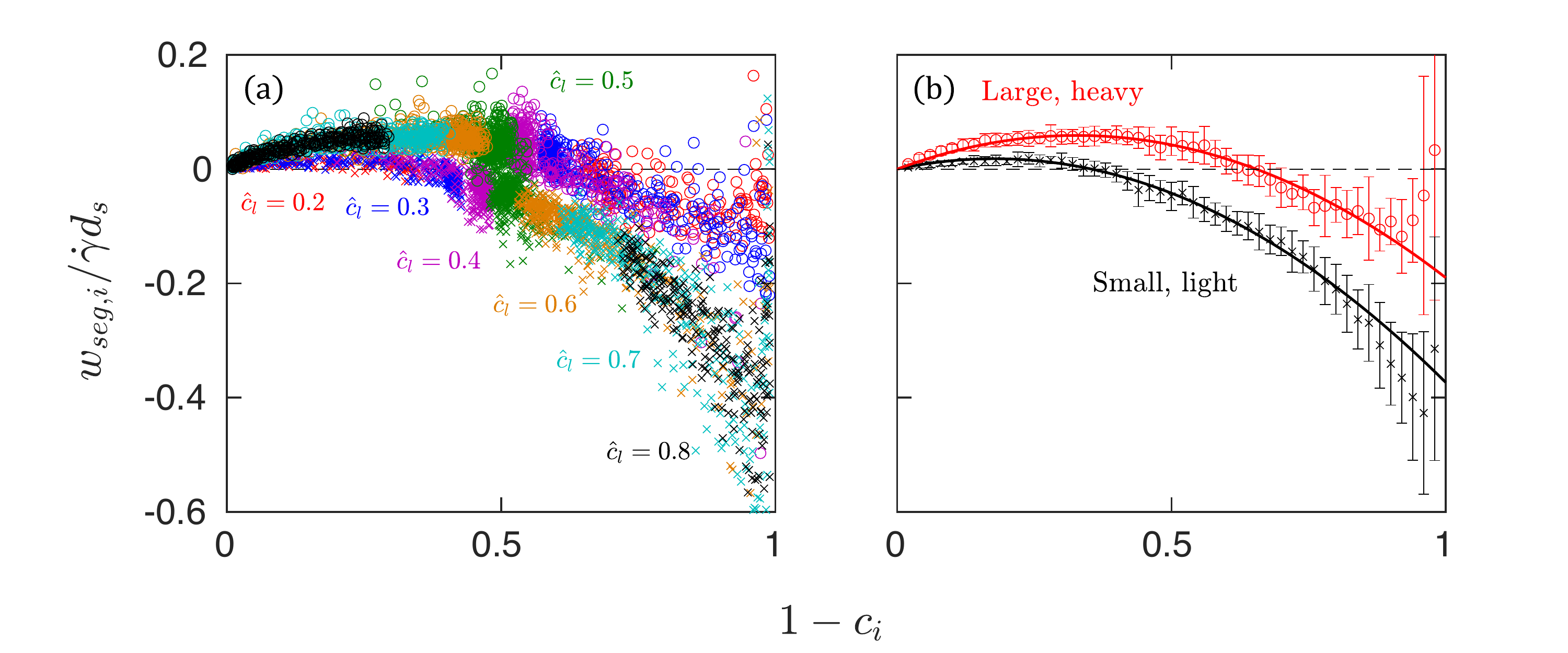}}
     \caption{Segregation velocity dependence of large, heavy ($\Circle$) and small, light ($\cross$) particles on local concentration of the other species, $1-c_i$, for $R_d=2$ and $R_\rho=4$. (a) Local segregation velocity for different feed concentrations $\hat c_l$ (symbol colours) calculated in each bin throughout the entire flowing layer averaged over 500 frames corresponding to 5$\,$s of simulated time.  (b) Data from (a) averaged over 0.02 increments of $1-c_i$ to reduce scatter and more clearly show the data trend. Error bars represent the standard deviation for each increment of $1-c_i$.
Solid curves are fits to equation~(\ref{quadkappa}).
}
         \label{wpdem}
\end{figure}

The local values of the velocities, concentrations, and diffusion coefficients shown in figure~\ref{contour} are used to determine the local segregation velocity using equation~(\ref{wpi}) for different values of the feed concentration of large particles, $\hat c_l$. An example of the resulting dimensionless segregation velocity data scaled by $d_s$ and shear rate, $\dot\gamma$, plotted against the concentration of the other species is shown in figure~\ref{wpdem} for $R_d=2$ and $R_\rho=4$.
Different colours in figure~\ref{wpdem}(a) represent simulations with different values for $\hat c_l$, which ensures a full range of local concentrations ($0.01<c_i<0.99$) in the plot. 
Because the data come from different depths and positions along the flowing layer of the bounded heap, a wide range of shear rates, concentrations, and segregation velocities are represented in the figure.
Although there is substantial scatter in the data due to the stochastic nature of granular flows, it is clear that there are two distinct curves represented in figure~\ref{wpdem}(a), an upper one for the LH particles and a lower one for the SL particles. It is also clear that $w_{seg,l}$ (upper set of data) changes from positive (upward segregation) to negative (downward segregation), depending on concentration. For $1-c_l<0.6$ [circles at the top left of figure~\ref{wpdem}(a)] $w_{seg,l}$ is positive, indicating that LH particles tend to rise in the flowing layer for small values of $1-c_l=c_s$; for $1-c_l>0.6$ [circles at the top right of figure~\ref{wpdem}(a)] $w_{seg,l}$ is negative, indicating that LH particles sink deeper in the flowing layer.
Results are analogous for SL particles, i.e. for small values of $1-c_s=c_l$ ($1-c_s<0.4$) SL particles rise, albeit at a smaller segregation velocity than LH particles, while for larger values of $1-c_s$, SL particles sink. 
The overlap between the data shown in figure~\ref{wpdem}(a) for separate simulations with seven different feed concentrations indicates the robustness of the results.

To more clearly show trends, figure~\ref{wpdem}(b) averages the data in figure~\ref{wpdem}(a) over 0.02 increments of $1-c_i$. 
Here the small positive segregation velocity for both species at small values of $1-c_i$ and the negative segregation velocity for both species at large values of $1-c_i$ is evident. 
The segregation velocity of both species is zero at $1-c_i=0$, which corresponds to the limit of no particles of the other species being present (monodisperse flow with no segregation possible).
However, the segregation velocity as $1-c_i$ approaches 1 is finite, as would be expected since this corresponds to a very low concentration of species $i$ amongst many particles of the other species (a single intruder particle in the limit). However, this low concentration leads to large variability in the measurement of $w_{seg,i}$ and correspondingly large error bars.

Returning to the expressions for the segregation velocity discussed in \S\ref{section1}, it is clear that the linear relation of equation~(\ref{wps}) is inappropriate for the data in figure~\ref{wpdem}, even though it works well under many other circumstances, particularly for mixtures of particles having similar concentrations \citep{schlick2015modeling,xiao2016modelling}. However, since previous studies have shown that the segregation velocity can be more accurately modelled using a quadratic polynomial, equation~(\ref{quadkappa}) \citep{gajjar2014asymmetric,van2015underlying,jones2018asymmetric}, particularly for values of $1-c_i$ near 0 or 1, we consider that form here.
Figure~\ref{wpdem}(b) shows that fits to equation~(\ref{quadkappa}) match well with the segregation velocity data. In particular, the fitted curves intersect the dashed horizontal line corresponding to $w_{seg,i}=0$ at $c_l\approx 0.35$ and $c_s\approx 0.65$. At these concentrations, which sum to 1 as they should, particles no longer segregate, apparently due to the balance between percolation and buoyancy.

\begin{figure}
    \centerline{\includegraphics[width=5.2 in]{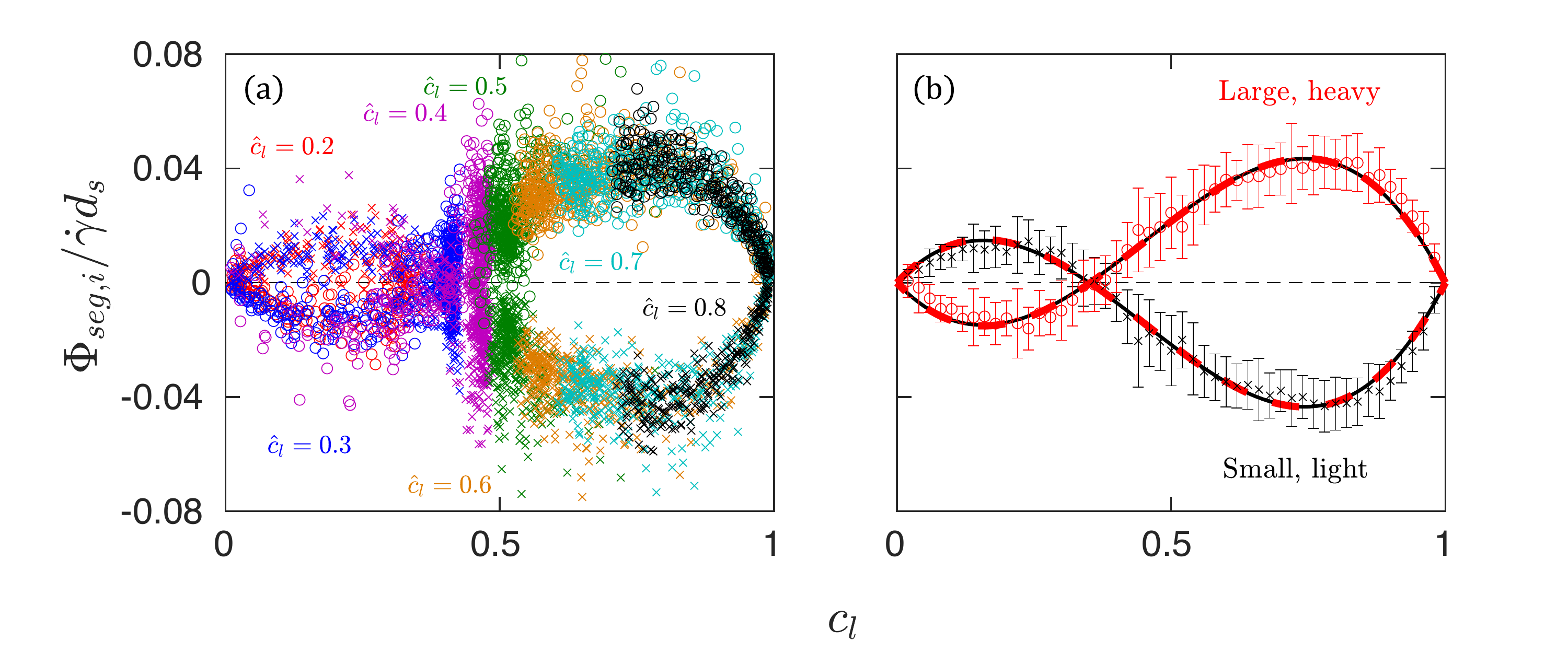}}
     \caption{Segregation flux dependence of large, heavy ($\Circle$) and small, light ($\cross$) particles on local concentration of large, heavy particles, $c_l$, for $R_d=2$ and $R_\rho=4$. (a) Local segregation flux for different feed concentrations $\hat c_l$ (symbol colours) calculated in each bin throughout the entire flowing layer and averaged over 500 frames corresponding to 5$\,$s of simulation time.  (b) Data from (a) averaged over 0.02 increments of $c_l$. 
     Error bars represent the standard deviation for each averaging interval of $c_l$.
Red dashed curves are quadratic fits of equation~(\ref{quadphi}) using only the data for large, heavy particles ($\textcolor{red}{\Circle}$) in the plot, while black solid curves (which appear dashed because they are behind the red dashed cruves) are fits to the data for small, light particles ($\textcolor{black}{\cross}$) in the plot.
}
         \label{flux}
\end{figure}

The measurement uncertainty for the segregation velocity increases with $1-c_i$ resulting in a mismatch between the data and the curve for large $1-c_i$ shown in figure~\ref{wpdem}.
To reduce uncertainty, particularly near the single particle intruder limit, the segregation velocity data in figure~\ref{wpdem} can be recast as the species segregation flux $\Phi_{seg,i}=w_{seg,i}c_i$.  As is evident for both the entire data set in figure~\ref{flux}(a) and the averaged data in figure~\ref{flux}(b), the two measured species segregation fluxes are always equal and opposite at any particular local value of the large particle concentration, $c_l$, as expected. It is also evident that the segregation flux direction reverses at $c_l=0.35$ where the segregation velocity in figure~\ref{wpdem} is zero.  The data in figure~\ref{flux} are well-fit by the quadratic-in-concentration segregation velocity given in equation (\ref{quadkappa}), which corresponds to a cubic in $c_i$ segregation flux:
\begin{equation}
\Phi_{seg,i}=d_s\dot\gamma c_i [A_{i}+B_{i}(1-c_i)](1-c_i),
\label{quadphi}
\end{equation}
where $A_i$ and $B_i$ are coefficients that we will show depend on both the size and density ratios. 
Equation~(\ref{quadphi}) automatically satisfies the requirement that the flux is 0 at $c_l=0$ and 1.

\begin{figure}
    \centerline{\includegraphics[width=5.2 in]{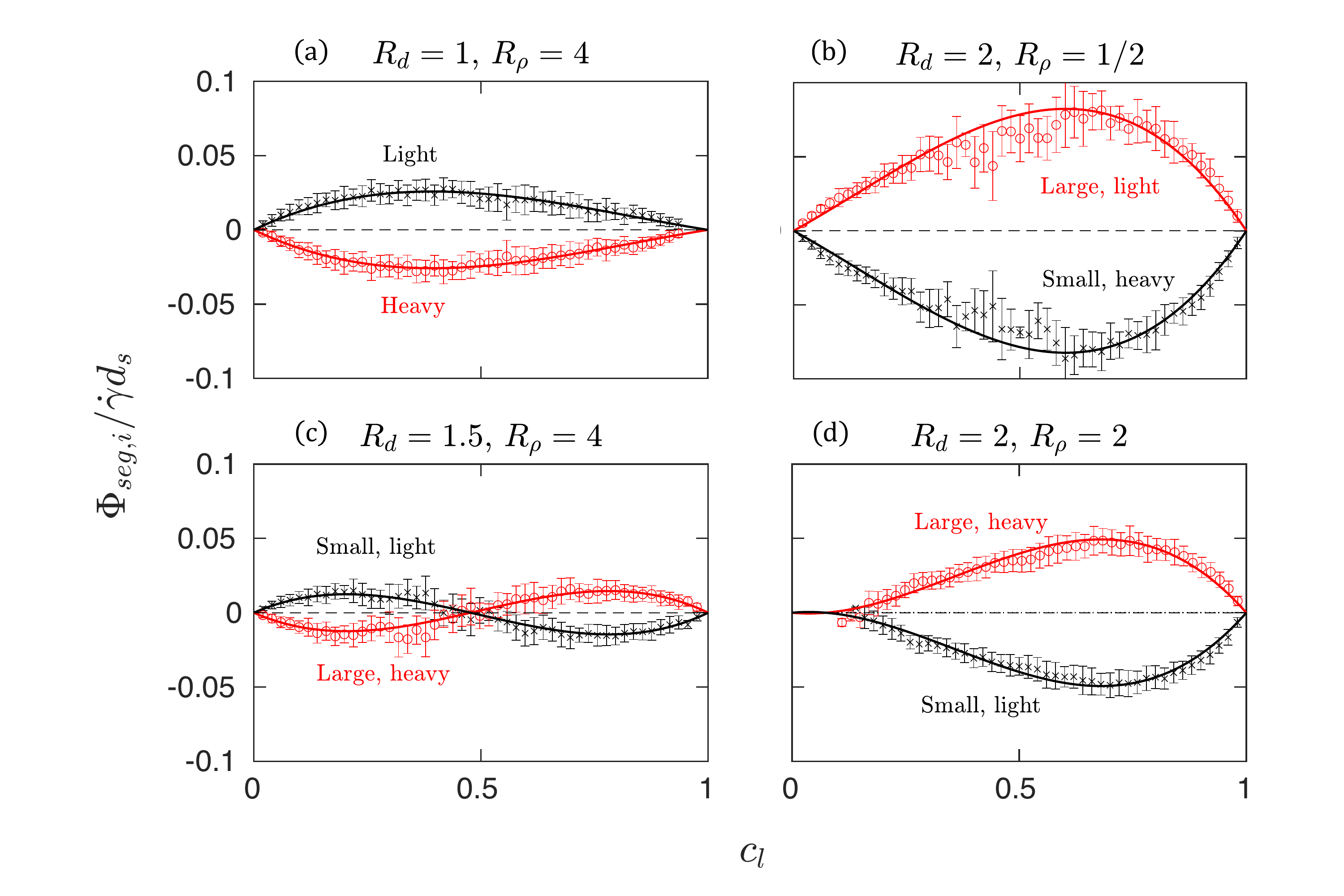}}
     \caption{ Segregation flux data for large ($\textcolor{red}{\Circle}$) and small particles ($\textcolor{black}{\cross}$) averaged over 0.02 increments of $c_l$ [subscript $l$ refers to heavy particles for D-system in (a)] for different size and density ratios. Solid curves are fits of equation~(\ref{quadphi}) to the data for large particles ($\textcolor{red}{\Circle}$). Error bars represent the standard deviation for each averaging interval of $c_l$.
     }
         \label{fourcase}
\end{figure}

\begin{figure}
    \centerline{\includegraphics[width=5.1 in]{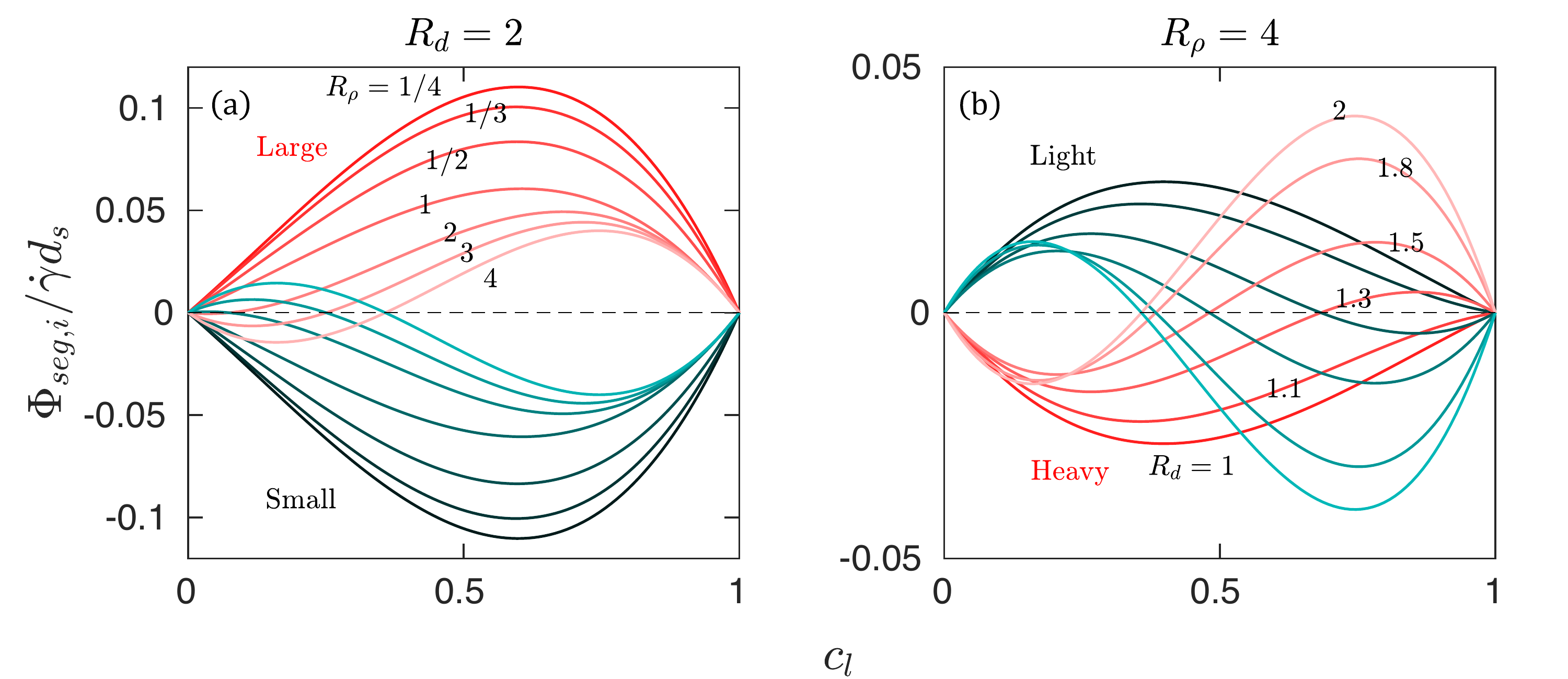}}
     \caption{Segregation flux vs. large particle concentration, $c_l$, from fits of equation~(\ref{quadphi}) to segregation flux data from simulations with (a) $R_d=2$ and various $R_\rho$ and (b) $R_\rho=4$ and various $R_d$.
     Red-green colour pairs represent simulations with different combinations of $R_d$ and $R_\rho$.
     Simulation data and error bars as plotted in figures~\ref{flux} and \ref{fourcase} are omitted for clarity.
     }
         \label{fluxit}
\end{figure}

Although equation~(\ref{quadphi}) can be made to fit the data in figure~\ref{flux} quite well, it is a phenomenological model that lacks a physical basis.
An alternative approach for considering the segregation is to solve the momentum equation supplied with species-specific stresses and an interspecies drag \citep{gray2005theory} such that the reversed segregation flux can be justified from the standpoint of force balance.
However, both approaches require empirical fits, either of the coefficients $A_i$ and $B_i$ in the segregation flux model [equation~(\ref{quadphi})] or through the force models related to species-specific stresses \citep{marks2012grainsize,tunuguntla2014mixture,tunuguntla2017comparing} or drag \citep{duan2020segregation} in momentum-based models. 
In fact, \citet{gray2015particle} assume a simple stress partitioning function and a linear interspecies drag and mathematically prove that the segregation can reverse direction for different values of concentration in SD-systems.
However, their model does not account for the asymmetric concentration dependence of size segregation \citep{gajjar2014asymmetric,van2015underlying,jones2018asymmetric}.
\textcolor{black}{In addition, their results indicate that the greatest segregation flux occurs for rising species at $c_i<0.5$ and for sinking species at $c_i>0.5$, which is opposite to the simulation results discussed below in reference to figures~\ref{fourcase} and \ref{fluxit}.}

Of the four empirical coefficients of the model described by equation~(\ref{quadphi}) [equivalently (\ref{quadkappa})], $A_l$, $B_l$ and $A_s$, $B_s$ (two for each species), only two are independent due to the constraint of volume conservation at all local concentrations, i.e. $\Phi_{seg,l}+\Phi_{seg,s}=0$.
For example, if $A_l$ and $B_l$ are determined for the large species, then the coefficients for the small species are
$A_s = -(A_l + B_l)$ and $B_s=B_l$.
Consequently, although there are two sets of data in figure~\ref{flux}(b), either curve fit can be used to determine the values of $A_i$ and $B_i$ for the other curve fit. For example, the segregation flux data for large particles in figure~\ref{flux}(b) (red circles) are used to find $A_l$ and $B_l$ first, and then $A_s$ and $B_s$ are calculated based on volume conservation. 
As shown in figure~\ref{flux}(b), the dashed red curves, which are based on the large particle data, also agree well with the small particle data.
Fitting equation~(\ref{quadphi}) to small particle flux instead of large particle flux results in nearly identical fits as is evident in figure~\ref{flux}(b) where the solid black curves (appearing as dashed because they are plotted behind the overlying red dashed curves) are fits to the small particle flux. 
That the segregation velocities for large and small particles are calculated independently in each bin in the DEM simulation, and that the curves for the segregation fluxes match not only the data but also each other, whether calculated from small or large particle data, demonstrates the robust nature of the approach for modelling the segregation [equations~(\ref{quadkappa}) and (\ref{quadphi})] and calculating $w_{seg,i}$, $c_i$, $\dot\gamma$, and $D$ as described in \S\ref{section2}.

Having shown that the segregation direction is concentration dependent for a specific pair of size and density ratios ($R_d=2$, $R_\rho=4$), we now illustrate how the segregation flux varies with concentration for different size and density ratios. Figure~\ref{fourcase} shows four different cases for four distinct ($R_d$, $R_\rho$) pairs.
The $R_d=1$ and $R_\rho=4$ case in figure~\ref{fourcase}(a) reduces to density only segregation, for which light particles rise and heavy particles sink.
Figure~\ref{fourcase}(a) also indicates that the heavy particle concentration is always less than 1 [no data for $c_l>0.94$ in figure~\ref{fourcase}(a); here subscript $l$ refers to heavy particles in D-systems] while the light particle concentration ($1-c_l$) reaches 1, consistent with previous studies on plane shear flows for D-systems \citep{fry2019diffusion}.
This is likely caused by asymmetric concentration dependence of density segregation in which a single heavy particle ($c_l\approx0$) can move through surrounding light particles but light particles get stuck among heavy particles at low concentration ($c_l\approx1$). 
Figure~\ref{fourcase}(b) shows an enhanced segregation case for $R_d=2$ and $R_\rho=1/2$, in which the particles quickly segregate and remain fully segregated for most of the flowing layer. 
The measurement uncertainties in figure~\ref{fourcase}(b) are larger than those of the other three cases, especially for $0.3<c_l<0.7$, because the strong segregation reduces the number of data points in this concentration range.

Unlike the two ``uni-directional" segregation cases in figures~\ref{fourcase}(a,b) in which percolation and buoyancy act in the same direction, percolation and buoyancy compete with each other for $R_d=1.5$ and $R_\rho=4$ in figure~\ref{fourcase}(c) and $R_d=2$ and $R_\rho=2$ in figure~\ref{fourcase}(d). In these cases, the segregation direction is concentration dependent. 
For $R_d=1.5$ and $R_\rho=4$ the overall segregation flux is small due to the near balance between the two segregation mechanisms. 
For $R_d=2$ and $R_\rho=2$ the concentration of LH particles at which segregation flux reverses is about 0.2, and the two segregation mechanisms are nearly balanced for $c_l\le0.2$, so particles remain mixed. As a result, simulation data are only available for $c_l\ge0.2$ in figure~\ref{fourcase}(d).  

Similar to figure~\ref{flux}(b), fits to equation~(\ref{quadphi}) accurately capture the segregation flux data for all four cases in figure~\ref{fourcase}. 
The applicability of the segregation model for various size and density ratios is demonstrated by the close agreement between the model predictions and the segregation flux data for not only opposing segregation mechanisms in figure~\ref{fourcase}(c,d) but also uni-directional segregation mechanisms in figure~\ref{fourcase}(b) or density only segregation in figure~\ref{fourcase}(a).

To illustrate how density ratio alone affects segregation in SD-systems, fits of equation~(\ref{quadphi}) to the simulation data for $R_d=2$ with $R_\rho$ varying from 1/4 to 4 are plotted in figure~\ref{fluxit}(a) (the simulation data and error bars are omitted for clarity).  
The segregation flux for the two species is always symmetric about $\Phi_{seg,i}=0$ due to volume conservation, and it increases for both species with decreasing $R_\rho$ at all concentrations as would be expected as the large particles become lighter.
For $R_\rho<1$ the segregation is uni-directional regardless of concentration because the particle size and density segregation mechanisms are in the same direction.
On the other hand, reversed segregation is possible when the two segregation mechanisms oppose each other for $R_\rho>1$, and in such cases the concentration at which the segregation flux reverses increases with $R_\rho$.
In addition, the segregation flux is nearly independent of density ratio as $c_l$ approaches 1 for $R_\rho\ge1$, indicating that buoyancy has little influence on segregation flux near the single SL intruder particle limit.

\begin{figure}
    \centerline{\includegraphics[width=2.6 in]{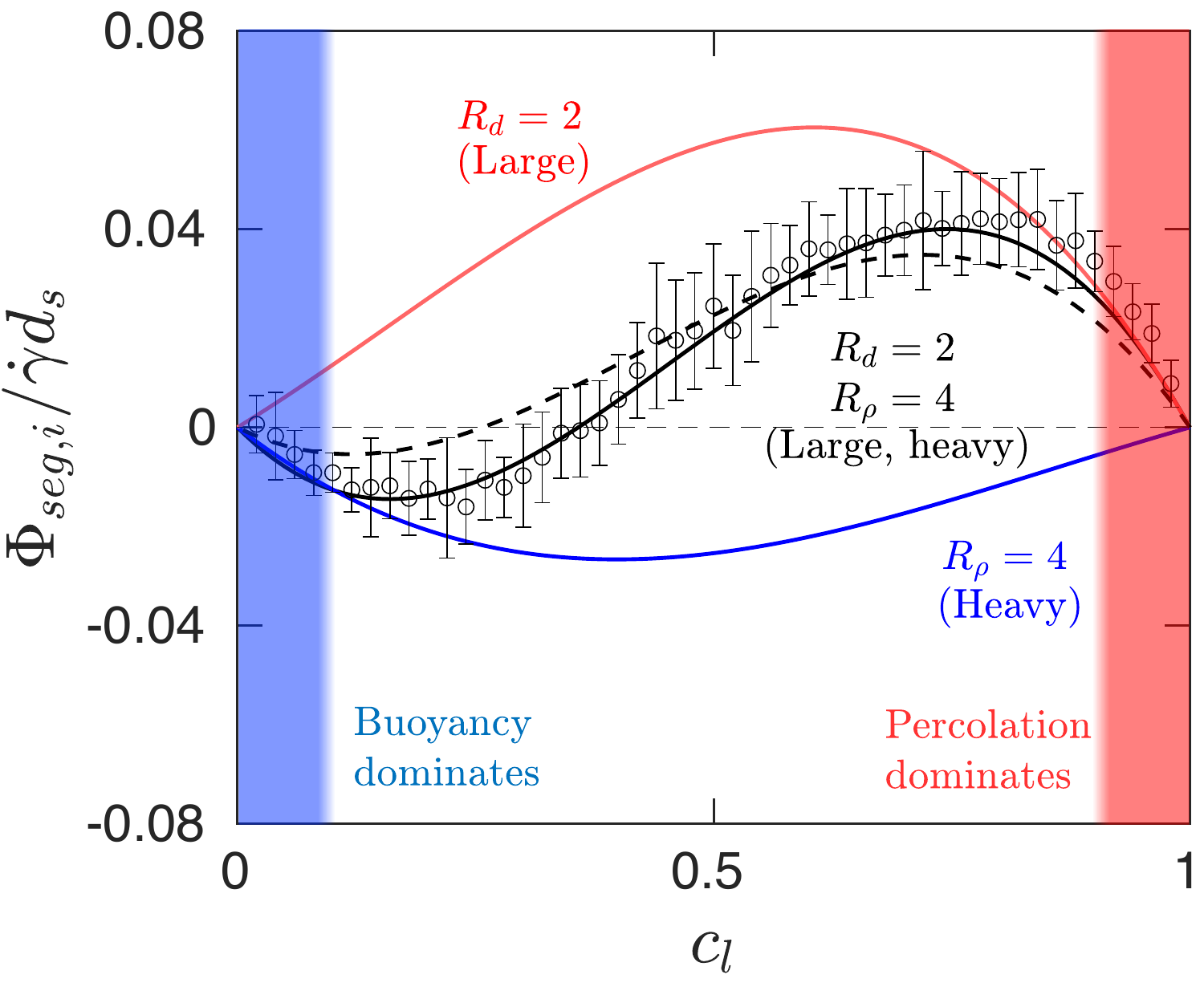}}
     \caption{Fits of equation~(\ref{quadphi}) to the segregation flux data for large particles in S-system ($\textcolor{red}{\text{red}}$), heavy particles in D-system ($\textcolor{blue}{\text{blue}}$), and large, heavy (LH) particles in SD-system ($\textcolor{black}{\text{black}}$). 
     Data for LH particles ($\Circle$) in SD-system are identical to those in figure~\ref{flux}(b).
     Dashed curve is the sum of segregation fluxes for S- and D-systems.  
     Blue and red coloured areas indicate the concentration ranges where buoyancy and percolation dominate, respectively.
     }
         \label{fluxcomb}
\end{figure}

Figure~\ref{fluxit}(b) plots similar results for constant density ratio $R_\rho=4$ and different values of the size ratio $R_d$, where the situation is more complicated. For similar size particles ($R_d$ near 1), the flux decreases as $R_d$ increases until the flux reverses between $R_d=1.1$ and $R_d=1.3$. Increasing $R_d$ further increases the magnitude of the flux so that by $R_d=2$ the upward flux of the LH particles is substantially larger than the downward flux of heavy particles of the same size ($R_d=1$).
The curves overlap for $c_l<0.1$, indicating that the segregation flux is nearly independent of size ratio (i.e., percolation is negligible in situations approaching the single LH particle limit).   
Similar behaviour is evident for the SL particles but in the opposite direction.

As expected, segregation goes to zero at the limits of $c_l=0$ and $c_l=1$.
One might hypothesize that the segregation behaviour of a SD-system at these two limits is similar to that of a S-system of the same size ratio or a D-system of the same density ratio.
To test this hypothesis for the segregation behaviour at the two concentration limits, figure~\ref{fluxcomb} plots segregation flux of large particles for a S-system of $R_d=2$, heavy particles for a D-system of $R_\rho=4$, and LH particles for a SD-system of $R_d=2$ and $R_\rho=4$.
Solid curves are fits to the simulation results and symbols represent the segregation flux data for LH particles. 
The segregation flux of LH particles at $c_l \lessapprox 0.2$ (blue coloured area) nearly equals that of heavy particles in the D-system having the same $R_\rho$, indicating that buoyancy dominates. 
Likewise, for values of $c_l\gtrapprox0.8$ (red coloured area) the segregation flux of LH particles matches that of large particles in the S-system having the same $R_d$, indicating percolation dominates at this limit. 
Between these two limits, for $0.2 \lessapprox c_l \lessapprox 0.8$, the LH curve deviates substantially from both the $R_d=2$ curve and the $R_\rho=4$ curve. 
One might speculate that the combined effect of size and density is simply a linear combination of the size effect and the density effect. 
To test this, consider the dashed curve which represents the sum of segregation fluxes for the S- and the D-systems. The mismatch between the data and the dashed curve clearly demonstrates that segregation model in an SD-system is not a simple linear combination of that for the corresponding S- and D-systems.

\begin{figure}
    \centerline{\includegraphics[width=5.1 in]{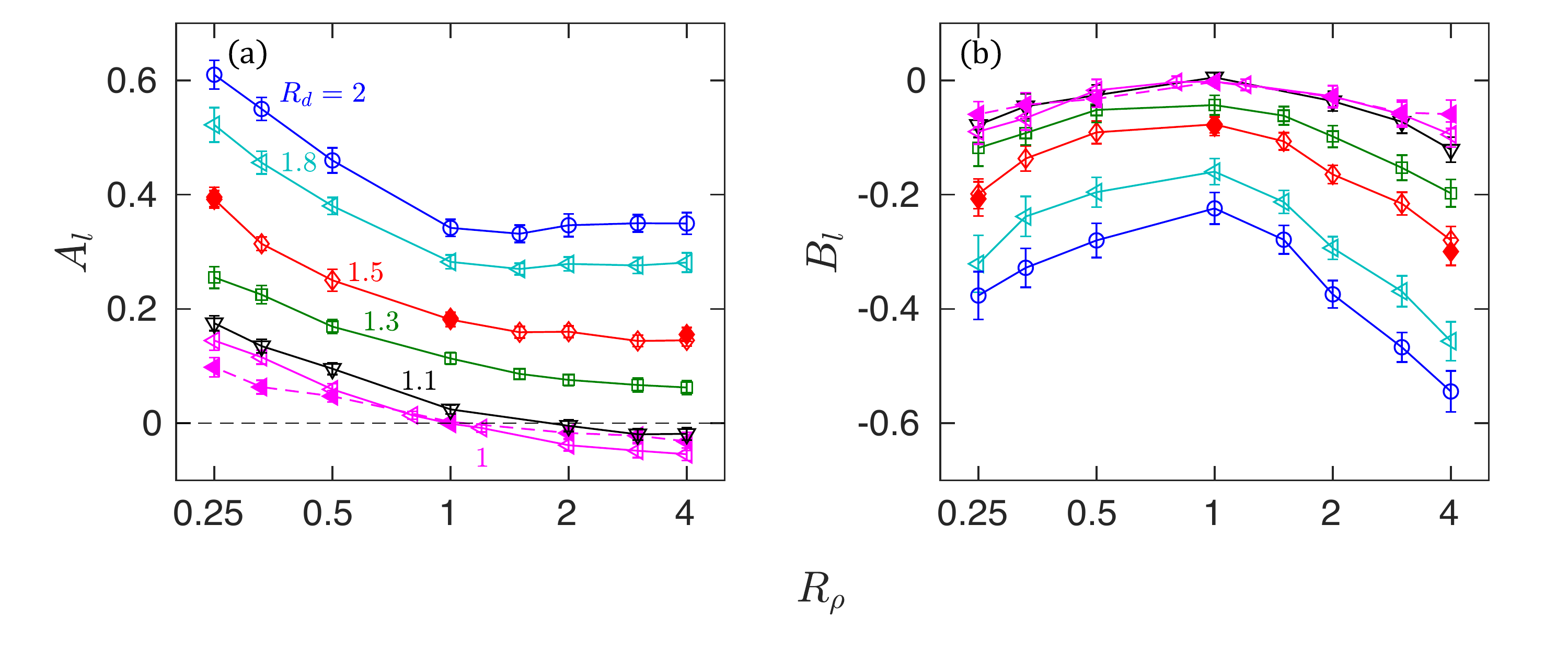}}
     \caption{
     Quadratic segregation velocity model coefficients: (a) $A_l$  and (b) $B_l$ vs. density ratio $R_\rho$ for various $R_d$ as indicated in (a). Filled symbols for $R_d=1$ and 1.5 represent simulations with no intra-species variation in particle diameter in contrast to simulations with 10\% size polydispersity (open symbols). Error bars represent 95\% confidence bounds based on the fit to the data.      }
      \label{coeff}
\end{figure}

Figures~\ref{wpdem}-\ref{fluxcomb}  show that the quadratic functional form of the segregation velocity in equation~(\ref{quadkappa}) and, equivalently, the cubic functional form of segregation flux in equation~(\ref{quadphi}) are generally applicable to different combinations of size and density ratios. The segregation behaviour can be predicted for arbitrary $R_d$ and $R_\rho$ combinations once the model coefficients $A_i$ and $B_i$ are known.
To more fully characterize the dependence of $A_i$ and $B_i$ on particle size and density ratios, 126 simulations were performed (three simulations per case) for $R_d=1$, 1.1, 1.3, 1.5, 1.8, 2 and $R_\rho=$ 1/4, 1/3, 1/2, 1, 2, 3, 4.
Here we focus on $R_d\le2$ because outside this range the segregation velocity in S-systems changes its dependence on $R_d$, suggesting the possibility of a change in the size segregation mechanism \citep{golick2009mixing,schlick2015modeling}. 
Furthermore, computational requirements increase substantially with increasing size ratio because of the increasing burden of contact detection with increasing $R_d$.
Figure~\ref{coeff} plots fitted values of the corresponding coefficients $A_l$ and $B_l$.
Independent of $R_\rho$, $A_l$ increases with $R_d$ as shown in figure~{\ref{coeff}}(a).
For fixed $R_d$, $A_l$ decreases with increasing $R_\rho$. 
Figure~\ref{coeff}(b) shows that $B_l$ becomes more negative with increasing $R_d$, and in all cases the magnitude of $B_l$ is smallest near $R_\rho=1$. 
It is tempting to try to collapse the data in figure~\ref{coeff}, but $A_l$ and $B_l$ are dimensionless parameters so they are not subject to scaling, and equation~(\ref{quadphi}) does not suggest any physical basis for collapsing the data.
As noted earlier, finding $A_s$ and $B_s$ using small particle data should be equivalent to finding $A_l$ and $B_l$ using large particle data due to the segregation flux of one being the negative of the segregation flux of the other.
Indeed, the values for $A_s$ and $B_s$ are within 5\% of those for $A_l$ and $B_l$ for all of the cases considered in figure~\ref{coeff}.

One further consideration is that for the simulations up to this point, each particle species has a $\pm$10\% variation in particle diameter, which is a common approach in DEM simulations to avoid ordered packing structures that can occur for identically sized particles. 
However, variation in particle size can create variations in segregation properties, particularly for $R_d$ close to 1 because the size distributions for the two species may overlap \citep{gao2020broad}. 
To assess the impact of a narrow range of particle sizes in combination with differences in particle densities, simulations are also conducted for particle species of uniform size (filled data points in figure~\ref{coeff} for $R_d=1$ and 1.5). 
As shown in figure~\ref{coeff}, the distribution of particle sizes has no effect on the resultant model coefficients for $R_d=1.5$. However, for $R_d=1$, the magnitude of $A_l$ for two particle species of uniform size is less than that for species with a narrow range of particle diameters, while $B_l$ remains nearly the same, indicating that the slight polydispersity promotes segregation for simulations where $R_\rho\ne 1$ as expected. 

\begin{figure}
    \centerline{\includegraphics[width=2.7 in]{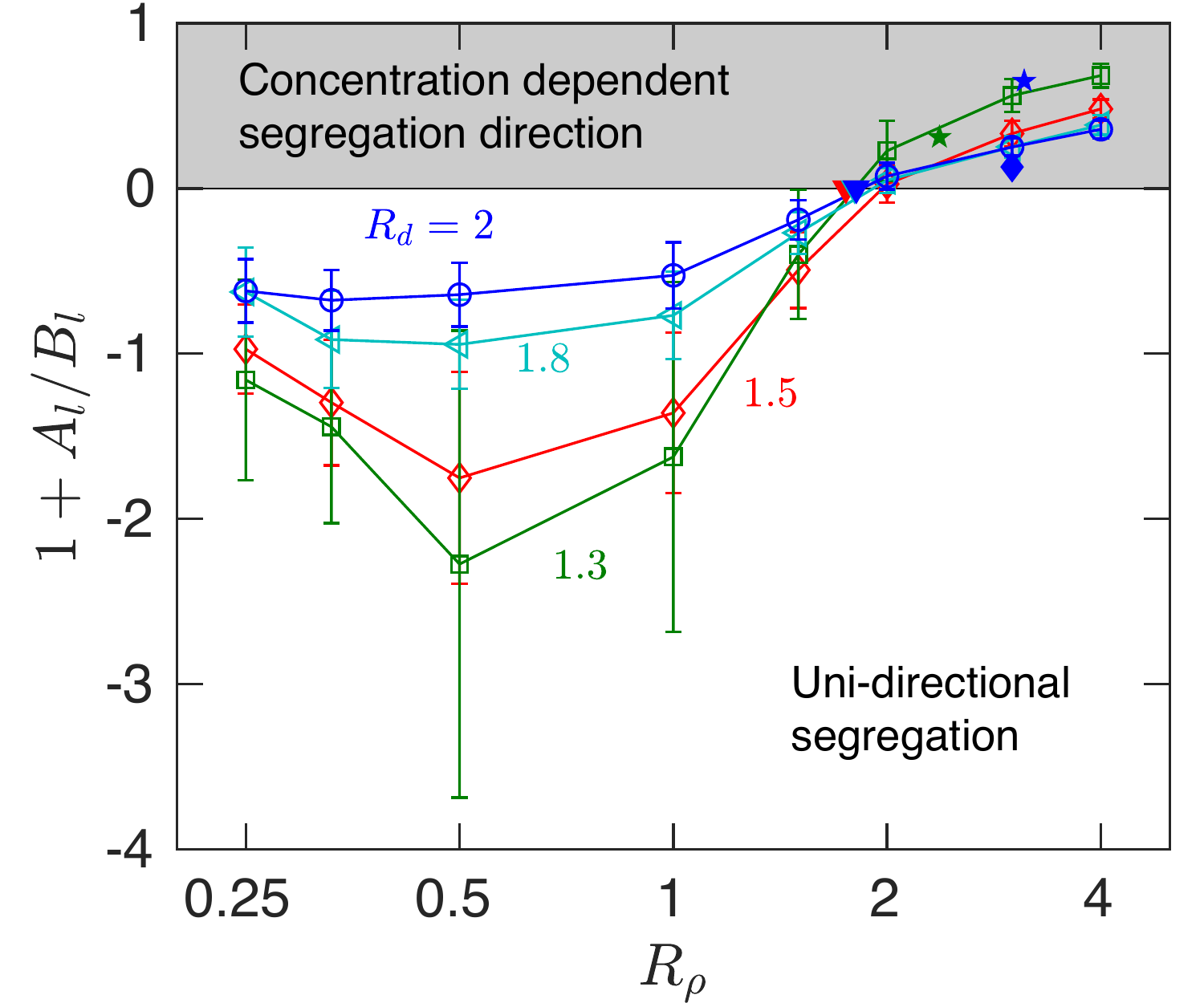}}
     \caption{
    \textcolor{black}{
    $1+A_l/B_l$ vs. density ratio $R_\rho$ for various $R_d$. Uncertainties are calculated as $\sqrt{(\frac{\Delta A}{A})^2+(\frac{\Delta B}{B})^2 }\frac{A}{B}$. For mixtures with $0<1+A_l/B_l<1$ (shaded area), segregation reverses at an equilibrium concentration $c_{l,eq}=1+A_l/B_l$ where $\Phi_{seg,l}=0$ [equation~(\ref{quadphi})]. Filled inverted triangles for $R_d\approx1.5$, 1.8 (hidden behind the blue triangle for $R_d\approx2$), 2 \citep{felix2004evidence}, and stars for $R_d\approx1.3$ and 2 \citep{alonso1991optimum} represent experimental results in rotating tumblers. Filled diamond is from heap flow experiments in appendix \ref{appendix}.
    }      
     }
      \label{coeff2}
\end{figure}

To more clearly convey the physical implications of the fitting parameters, consider that the segregation flux model in equation~(\ref{quadphi}) is cubic with fixed zeros (roots) at $c_l=0$ and $c_l=1$ independent of $A_l$ and $B_l$.  A third zero exists at $c_l=1+A_l/B_l$  which is physical when $0\le1+A_l/B_l\le1$.
Figure~{\ref{coeff2}} plots $1+A_l/B_l$ as a function of $R_\rho$ for all cases except $R_d=1$, 1.1, because those results are affected by size dispersity of the two species.
For mixtures with $1+A_l/B_l\le0$, segregation flux of large particles, $\Phi_{seg,l}$ from equation~(\ref{quadphi}), is always greater than zero regardless of species concentration, indicating that the segregation is uni-directional. However, for $0<1+A_l/B_l<1$, the segregation direction is concentration dependent, with the segregation direction reversing at an equilibrium concentration, $c_{l,eq}=1+A_l/B_l$.
It is evident that uni-directional segregation always occurs for $R_\rho\lessapprox 1.8$, while the direction of segregation is concentration dependent for $R_\rho \gtrapprox 1.8$.

To validate this claim, the rise-sink transition predicted in figure~{\ref{coeff2}} is compared to free-surface flow experiments in a rotating tumbler in which LH tracer particles ($c_l\approx 0$) with $R_d\approx$ 1.5, 1.8, and 2.0 maintain an intermediate radial position in the flow for $R_\rho\approx 1.8$ \citep{felix2004evidence}. This corresponds to the size and density ratios where curves for $R_d\approx$ 1.5, 1.8, and 2.0 intersect the $1+A_l/B_l=0$ ($c_{l,eq}=0$) horizontal line in figure~\ref{coeff2}. All three of these experimental data points for a rotating tumbler (inverted triangles, one of which is hidden behind the others) are indeed on the $1+A_l/B_l$ horizontal line at $R_\rho\approx1.8$, thereby confirming the predicted transition from uni-directional segregation to a concentration dependent segregation direction.  
Similar experiments for mixtures of large glass and small plastic particles in a tumbler with $R_d\approx1.3$ and $R_\rho\approx 2.4$ \citep{alonso1991optimum} show that the segregation direction changes at $c_{l,eq}=0.31$, and, again, a close match between the data point (green star) and the curve (green) is observed.
However, a mixture of large steel and small glass particles with $R_d\approx 2.3$ and $R_\rho\approx 3.1$ \citep{alonso1991optimum} has an equilibrium concentration at which the segregation direction changes of $c_{l,eq}=0.65$ (blue star), which is significantly larger than the predicted value ($c_{l,eq}\approx0.25$) for $R_d=2$. To further test the model for this particular combination of particle size and density ratios, a few simple experiments with similar large steel and small glass particles ($R_d\approx2$ and $R_\rho\approx3$) are conducted in a bounded heap flow (see appendix \ref{appendix}) similar to our previous experiments \citep{fan2012stratification}.  Based on these experiments, it is clear that the concentration at which segregation changes direction is between 0.11 and 0.26 (blue diamond), which is consistent with the predicted value of 0.25.

A practical concern for many industrial situations is assuring that a mixture of two species of particles remains mixed. Figure~\ref{fluxit} shows that mixtures of particles having certain combinations of size and density ratios have a segregation flux of zero at the equilibrium concentration and therefore remain mixed.  The question is how to use this information to intentionally design the particles making up the mixture to minimize segregation and remain mixed? In many practical situations the material of each particle species is fixed, thereby fixing the density ratio.  Likewise, the relative concentrations of the two species is typically specified to maintain certain properties of the overall mixture.  However, frequently the sizes of the particles of each species can be altered as desired.  This opens the possibility of specifying particle sizes to avoid segregation at a certain mixture concentration.

\begin{figure}
    \centerline{\includegraphics[width=3.2 in]{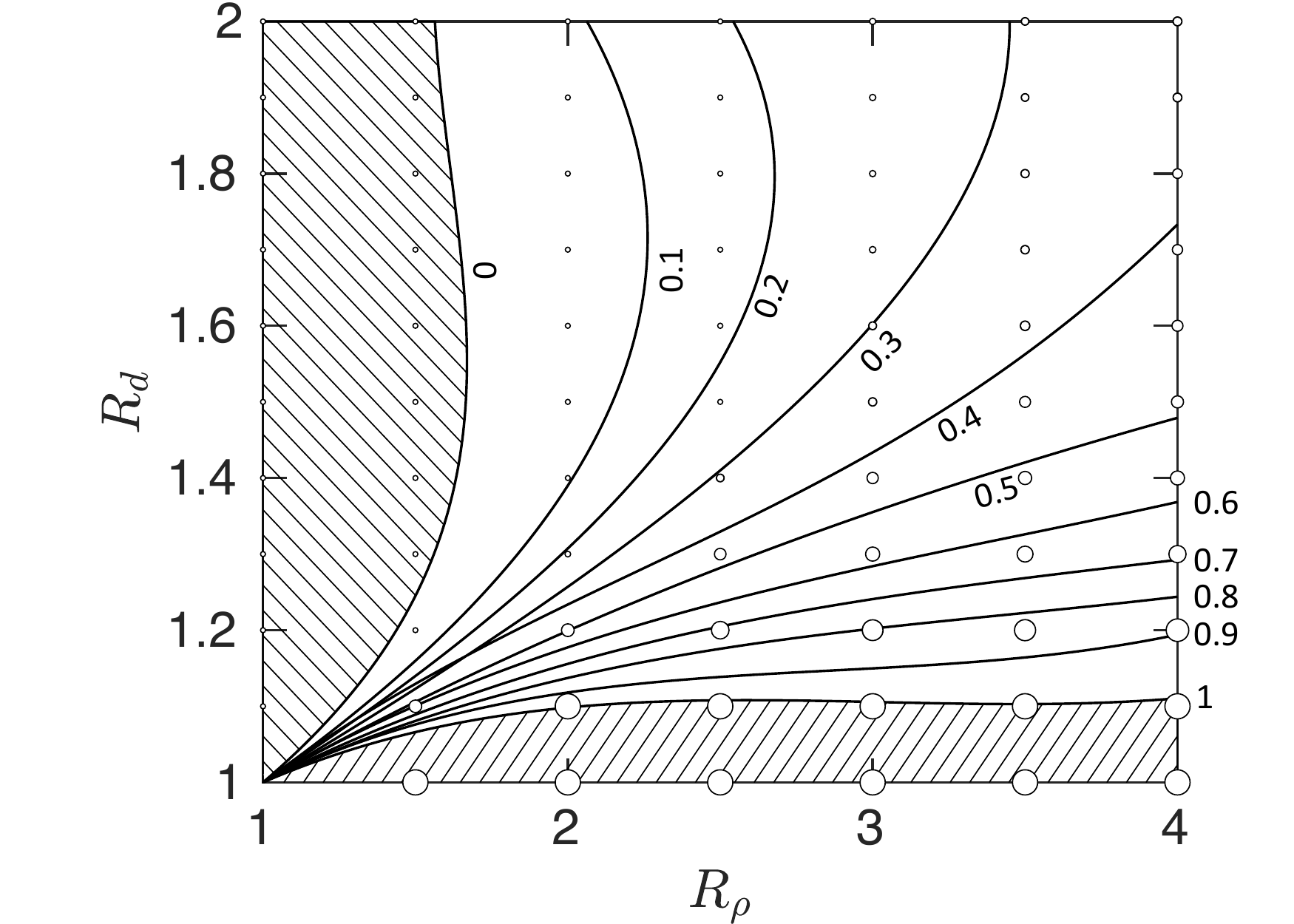}}
     \caption{
    \textcolor{black}{
    Equilibrium (no segregation) concentration of large particles $c_{l,eq}$ as a function of particle size and density ratios. Symbol ($\Circle$) diameter is proportional to $c_{l,eq}$ in the range 0 to 1. Iso-concentration curves for values of $c_{l,eq}$ are interpolated between data points. Particles stay mixed along the curve for $c_l=c_{l,eq}$ for the corresponding $R_d$ and $R_\rho$. Large particles rise if $c_l$ is greater than $c_{l,eq}$, whereas if $c_l$ is less than $c_{l,eq}$ large particles sink.   
Segregation is uni-directional (hatched regions) for $c_{l,eq}=0$ (large particles rise) and $c_{l,eq}=1$ (large particles sink).
    }      
     }
      \label{phase}
\end{figure}

The dependence of the equilibrium concentration $c_{l,eq}$ on particle size and density ratios is shown in figure~\ref{phase}, where iso-concentration curves for values of $c_{l,eq}=1+A_l/B_l$ are interpolated based on values of $A_l$ and $B_l$ for 77 different combinations of size and density ratios. 
The curves are cubic fits to the interpolated data forced to pass through ($R_\rho=1$, $R_d=1$), where segregation is necessarily zero.
Particles remain mixed along the curve for $c_l=c_{l,eq}$ for the corresponding $R_d$ and $R_\rho$.
Large particles rise if $c_l$ is greater than $c_{l,eq}$, whereas if $c_l$ is less than $c_{l,eq}$ large particles sink.
Segregation is uni-directional for $R_d$ and $R_\rho$ combinations in the hatched regions of figure~\ref{phase}. 
For $R_d$ and $R_\rho$ combinations to the left of $c_{l,eq}=0$, large particles rise, while large particles sink for combinations below $c_{l,eq}=1$, regardless of the mixture concentration.

To demonstrate the practical use of figure~\ref{phase}, consider a 20-80 mixture of large steel and small glass particles ($c_{l}=0.2$) with $R_d=2$ and $R_\rho=3$. The steel particles sink because the mixture concentration is less than the interpolated equilibrium concentration of $c_{l,eq}\approx 0.25$ at this combination of $R_d$ and $R_\rho$.
On the other hand, large steel particles would rise if their concentration is greater than $c_{l,eq}\approx 0.25$.

Perhaps more importantly, figure~\ref{phase} can be used to intentionally design a particle system so that the particles remain mixed by specifying the optimal size ratio $R_d$ to minimize segregation (i.e., optimize the particles remaining mixed) for any given density ratio $R_\rho$ and desired mixture concentration.
For example, a 50-50 mixture ($c_{l}=0.5$) of  small glass and large steel particles, corresponding to a density ratio $R_\rho= 3$, requires a size ratio of $R_d\approx 1.35$ to minimize segregation according to figure~\ref{phase}.

Note that for low concentrations of LH particles ($c_l<0.3$), the segregation behaviour is more complicated than at higher concentrations as the iso-concentration curves are nonmonotonic.
For example, to minimize segregation for a mixture with $R_\rho=2$ and $c_l=0.1$, there are two size ratios (i.e., $R_d\approx1.4$ and 2) for which the particles remain mixed.
This is consistent with previous studies on intruder particles under uniform shear, in which the lift-like force varies non-monotonically with size ratio having a maximum value near $R_d=2$ \citep{jing2020rising}.

\section{Continuum modelling of segregation}
\label{section4}

Previous studies show that a segregation velocity model combined with a transport equation can predict the segregation of an S- or D-system in quantitative agreement with experiments and simulations for different flow geometries \citep{fan2014modelling,schlick2015granular,xiao2016modelling,xiao2019continuum,isnergranular}. With the quadratic segregation velocity model presented here [equations~(\ref{quadkappa}) and (\ref{quadphi})], we extend this approach to SD-systems and compare the model predictions with simulation results for a range of bidisperse heap flows. This also further validates the coefficient values $A_i$ and $B_i$ determined in \S\ref{section3}, since the model predictions are compared to different flows than those from which $A_i$ and $B_i$ are determined.   

Similar to previous studies \citep{fan2014modelling}, we assume that flow kinematics and segregation are uncoupled and apply a modified advection-diffusion equation to predict the segregation of bidisperse granular heap flows like that shown in figure~\ref{scheme}.
Rearranging equation~(\ref{transport1}), utilizing conservation of volume, and noting that diffusion is significant for segregation only in the normal direction \citep{fan2014modelling}, results in
\begin{equation}
\frac{\partial c_i}{\partial t} + u \frac{\partial{c_i}}{\partial x}+w\frac{\partial c_i}{\partial z}+\partial\frac{w_{seg,i}c_i}{\partial z}=\frac{\partial}{\partial z} \Big( D\frac{\partial c_i}{\partial z} \Big).
\label{transport5}
\end{equation}
The transient term ${\partial c_i}/{\partial t}$ is zero in a reference frame rising with the surface for the steady state heap flow segregation considered here, and the segregation velocity $w_{seg,i}$ is given by equation~(\ref{quadkappa}).
Previous studies of quasi-2D heap flow kinematics \citep{fan2013kinematics} show that the streamwise velocity in the flowing layer is well approximated by
\begin{equation}
u(x,z)=\frac{kq}{\delta (1-e^{-k})} \Big( 1-\frac{x}{L}  \Big) e^{kz/\delta}.
\label{u}
\end{equation}
Here $k$ is a scaling constant set to $\ln(10)\approx 2.3$, so that $u(x,-\delta)=0.1u(x,0)$, consistent with the definition for $\delta$ earlier in this paper. 
Based on (\ref{u}) and the continuity equation the normal velocity is \citep{fan2013kinematics}
\begin{equation}
w(z)=\frac{q}{L(1-e^{-k})} (e^{kz/\delta}-1).
\label{w}
\end{equation}

Since equations (\ref{u}) and (\ref{w}) have been validated by experiments and simulations for size \citep{fan2013kinematics} and density \citep{xiao2016modelling} bidisperse granular materials in quasi-2D bounded heap, these functional forms should be readily extended to particle mixtures that differ in both size and density. 
We demonstrate the validity of this assumption for an example SD-disperse mixture with $R_d=2$, $R_\rho=4$, and $\hat c_l=0.5$ (see Supplementary Material). 

In dense granular flows, experimental \citep{bridgwater1980self,utter2004self} and computational \citep{fan2015shear,cai2019diffusion,fry2019diffusion} studies indicate that the diffusion coefficient, $D$, is proportional to the product of the local shear rate and the square of the local mean particle diameter,
\begin{equation}
D=C_{diff} \dot\gamma \bar d^{\,2},
\label{diffusion}
\end{equation}
where $\bar d=\sum c_id_i$ and $C_{diff}$ is a constant. 
Values for $C_{diff}$ have been reported in the range 0.01 to 0.1 \citep{hsiau1999fluctuations,savage1993studies,utter2004self,fan2014modelling,fan2015shear,fry2019diffusion,cai2019diffusion}. 
In this study, $C_{diff}$ is approximated as a constant equal to 0.046 based on the diffusion coefficient data measured from the heap flow simulations using equation~(\ref{MSD}).
This value is also consistent with a previous study on uniform shear flows, where $C_{diff}= 0.042$ \citep{fry2019diffusion}.
We further note that although the diffusion coefficient plays a role in the accurate determination of the segregation velocity (\ref{wpi}), the continuum segregation model is insensitive to the precise value \citep{fan2014modelling}.
The local shear rate $\dot\gamma$ in (\ref{diffusion}) is calculated from the velocity profile in (\ref{u}),  
\begin{equation}
\dot\gamma=\frac{\partial u}{\partial z}=\frac{k^2q}{\delta^2 (1-e^{-k})} \Big( 1-\frac{x}{L}  \Big) e^{kz/\delta}.
\label{gamma}
\end{equation}

With the bulk velocity $\pmb u$, the segregation velocity $w_{seg,i}$, and the diffusion coefficient $D$ determined, the continuum segregation model of (\ref{transport5}) can be used to obtain the local concentration of each species in the flowing layer.
Similar to previous studies \citep{fan2014modelling,schlick2015modeling,xiao2016modelling}, (\ref{transport5}) is non-dimensionalized using
\begin{equation}
\tilde x=\frac{x}{L},~~\tilde z=\frac{z}{\delta},~~\tilde u=\frac{u}{2q/\delta},~~\text{and}~\tilde w=\frac{w}{2q/L}.
\end{equation}
By substituting the segregation velocity (\ref{quadkappa}) and the diffusion coefficient (\ref{diffusion}) into (\ref{transport5}), the non-dimensional transport equation during steady filling takes the form
\begin{equation}
\tilde u \frac{\partial c_i}{\partial \tilde x}+\tilde w  \frac{\partial c_i}{\partial \tilde z} +  \frac{\partial }{\partial \tilde z} \bigg\{ \tilde{\dot\gamma} \big [A_i+B_i(1-c_i) \big](1-c_i) c_i \bigg\} = \frac {\partial}{\partial \tilde z}\bigg( C_{diff}\tilde {\dot\gamma} \frac{\bar d^2}{d_s\delta}  \frac{\partial c_i}{\partial \tilde z} \bigg),
\label{nond}
\end{equation}
where $\tilde {\dot\gamma}={Ld_s\dot\gamma/2q}$ is the non-dimensional shear rate.
Note that the local mean particle diameter $\bar{d}$ depends on local concentration, $c_i$.

Equation~(\ref{nond}) is solved numerically as an initial-boundary value problem using the $pdepe$ solver in MATLAB with global parameters matching the simulation input ($d_s$, $d_l$, $q$, $\hat c_l$) and measured directly from the simulations results ($\delta$, $L$) \citep{xiao2016modelling,deng2018continuum}. 
For accurate model predictions, the boundary conditions for (\ref{nond}) need to match those in the DEM simulations.
Following previous studies \citep{fan2014modelling,xiao2016modelling}, a well-mixed inlet boundary condition with local concentration independent of depth at the upstream end of the flowing layer is assumed, i.e. $c_{i} ({\tilde x=0,\tilde z})=\hat c_i$. 
At the top and bottom boundaries of the flowing layer, the segregation flux equals the diffusive flux, which are identical to those for size (or density) only segregation \citep{gray2006particle}, and is expressed as
\begin{equation}
 \big [A_i+B_i(1-c_i) \big](1-c_i) c_i=C_{diff} \frac{\bar d^2}{d_s\delta}  \frac{\partial c_i}{\partial \tilde z} ~~(\tilde z=0,~-1).
\label{bc}
\end{equation}
No boundary condition is needed at $\tilde x=1$ since both segregation and diffusion act in the $z-$direction and the streamwise velocity is zero.

\begin{figure}
    \centerline{\includegraphics[width=5.6 in]{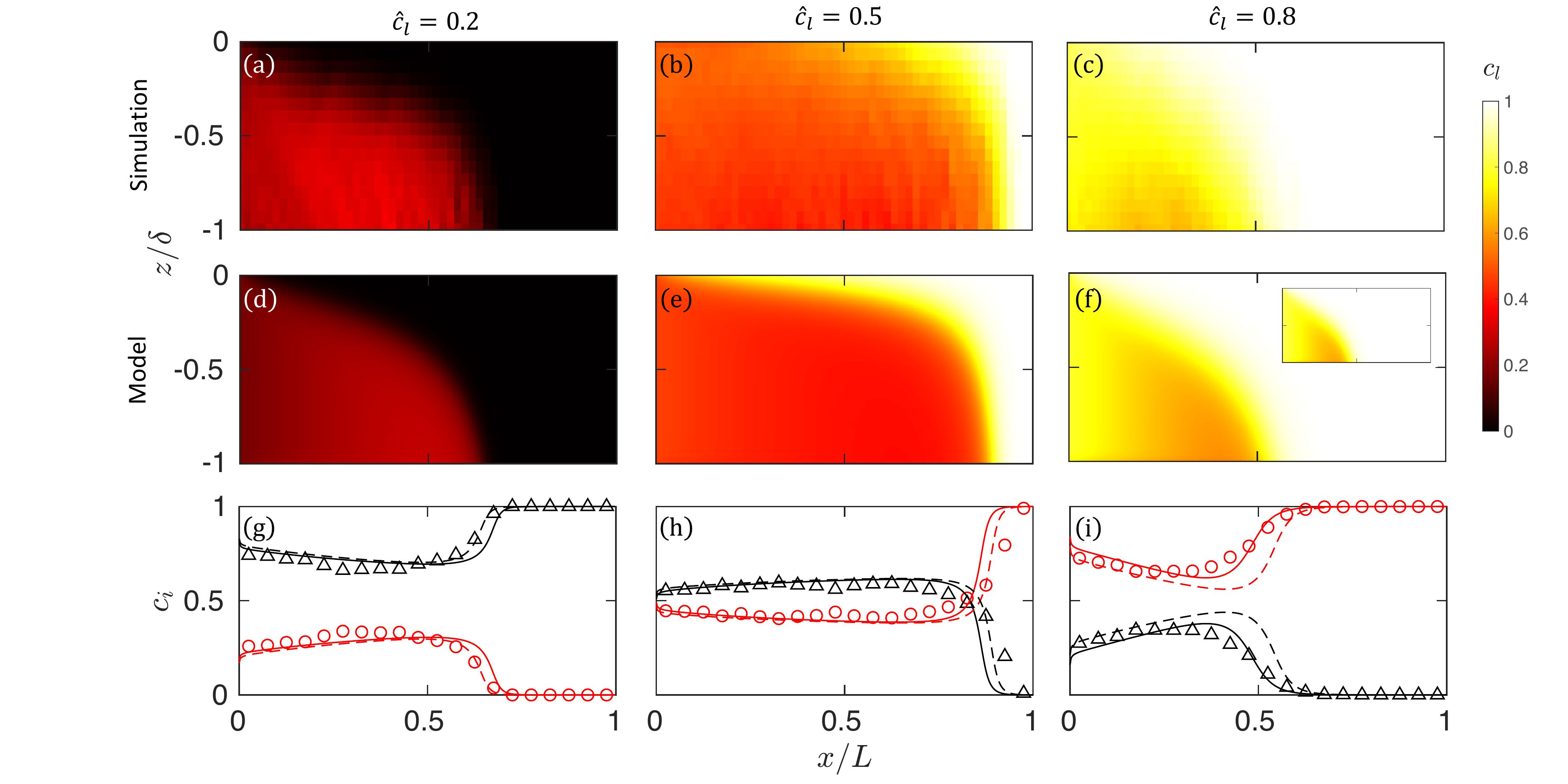}}
    \caption{
    Large, heavy (LH) particle concentration for different feed concentrations: $\hat c_l= 0.2$ (left column), $\hat c_l= 0.5$ (middle column), and $\hat c_l= 0.8$ (right column). 
    $c_l$ from (a-c) simulation and (d-f) continuum model with inlet boundary condition $c_{i}({\tilde x=0,\tilde z})=\hat c_i$. (g-i) $c_l$ for LH particles ($\textcolor{red}{\Circle}$) and $c_s$ for SL particles ($\textcolor{black}{\triangle}$) from simulation compared to model predictions using equation~(\ref{nond}) at the bottom of the flowing layer (curves), (i.e., $z/\delta=-1$ or $\tilde z=-1$) vs. streamwise position, $x/L$ for inlet boundary condition set to feed concentration $c_{i}({\tilde x=0,\tilde z})=\hat c_i$ (dashed) or mean concentration of particles deposited on the heap outside the feed-zone $c_{i} (\tilde x=0,\tilde z)=\langle c_{i}(\tilde x,\tilde z=-1)\rangle$ (solid). Inset: concentration fields corresponding to the model predictions of solid curve in (i). 
$R_d=2$, $R_\rho=4$, $q=20\,$cm$^2/$s, $\delta\approx1.5\,$cm. and $L=52\,$cm.     }
    \label{result}
\end{figure}

The predicted concentration profiles using a well-mixed inlet boundary concentration, $c_{i}({\tilde x=0,\tilde z})=\hat c_i$, are compared to the simulation results for $R_d=2$ and $R_\rho=4$ for three different inlet concentrations in figure~\ref{result} (corresponding to the simulations for the three inlet concentration examples shown in figure~\ref{comparison}).
Each of the panels in figure~\ref{result}(a-f) depicts the local concentration of large particles, $c_l$, in flowing layer corresponding to the rectangular box in figure~\ref{scheme} (inlet is the left edge of each panel, free surface is the top edge, bottom of the flowing layer where particles are deposited on the heap is the bottom edge, and downstream endwall is the right edge).
The continuum model results for the large particle concentration throughout the flowing layer shown in the second row of figure~\ref{result} are nearly identical to that for DEM simulation in the first row, demonstrating how the model matches simulation results.
Most significantly, the continuum segregation model captures the reversed segregation for different feed concentrations.
For $\hat c_l= 0.2$ [figure~\ref{result}(a,d)] LH particles quickly segregate to the bottom of the flowing layer in the upstream portion of the heap, while SL particles rise to the surface of the flowing layer and are advected further downstream.  
Increasing $\hat c_l$ to 0.5 reverses the segregation direction, such that LH particles rise toward the free surface and are advected downstream, as shown in figure~\ref{result}(b,e). Further increasing $\hat c_l$ to 0.8 results in an enlarged LH particle region in figure~\ref{result}(c,f) with SL particles only in the lower part of the upstream portion of the flowing layer.

Figures~\ref{result}(g-i) compare the streamwise concentration profiles from the simulations (data points) and continuum model predictions (dashed curves) at the bottom of the flowing layer ($z/\delta=-1$), which corresponds to the concentration of particles deposited on the heap. 
Again the reversal of the direction of the segregation is evident. 
For $\hat c_l=0.2$ the concentration of LH particles is highest for $x/L<0.6$ as they deposit on the upstream portion of the heap along with SL particles. Pure SL particles deposit on the downstream end of the heap. 
The situation reverses for $\hat c_l=0.5$ and 0.8 where SL particles deposit on the upstream portion of the heap along with a lower concentration of LH particles, and pure LH particles deposit on the downstream portion.
Moreover, that the continuum segregation model can predict the combined size and density segregation for different inlet concentrations as shown in figure~\ref{result} further validates the quadratic functional form of the segregation velocity in equation~(\ref{quadkappa}).

The difference between the continuum model predictions (dashed curves) and the simulations (data points) is small (less than 7\% on average) for $\hat c_l=0.2$ and 0.5. 
The slight deviations between simulation and model predictions in these two cases could be the result of many factors including bouncing particles near the free surface and the decreasing flowing layer thickness near the bounding endwall, neither of which are included in the model.  
In contrast, the discrepancy between the continuum model prediction and simulation results for $\hat c_l=0.8$ is relatively large in the middle portion of the heap, see figure~\ref{result}(i). 
In addition, it is evident in figure~\ref{result}(i) that the large particle concentration integrated over the length of the heap is larger for the simulations than for the continuum model and the small particle concentration is smaller.
This is a result of strong segregation combined with deposition in the feed-zone for $\hat c_l=0.8$, noting that the segregation flux is maximum near $c_l=0.8$ in figure~\ref{flux} for these particular values of $R_d$ and $R_\rho$.
Thus, small particles segregate to the bottom of the flow in the feed-zone to deposit on the heap before they start to flow down the heap.
As a result, the inlet of the flowing layer has a higher large particle concentration and a lower small particle concentration than the feed concentration (i.e., mixture falling onto the heap). 

To account for segregation in the feed-zone, we use a corrected inlet concentration for the model, $c_{i} (\tilde x=0,\tilde z)=\langle c_{i}(\tilde x,\tilde z=-1)\rangle$, which is the mean concentration at the bottom of the flowing layer outside of the feed-zone as determined from the simulation.
Alternatively, the mean inlet particle concentration at the inlet boundary where particles enter the flowing layer [$\langle c_{i}(\tilde x=0, \tilde z)\rangle$] could be used. 
However, because the rise velocity varies slightly near the feed-zone, particles deposited on the bed through the bottom of the flowing layer more accurately reflect the mean particle concentration entering the upstream end of the flowing layer in the simulations.
The corrected inlet concentrations measured from the three simulations in figure~\ref{result} are $\langle c_{i}(\tilde x,\tilde z=-1)\rangle=$ 0.19, 0.49, and 0.83 rather than the actual values of feed concentration of $\hat c_l=$ 0.17, 0.47, and 0.79, respectively.

Model predictions using the corrected inlet concentration (solid curves) match the simulation results more closely for $\hat c_l=0.8$ in figure~\ref{result}(f,inset) and (i), significantly reducing the discrepancy for $0.3<x/L<0.6$.
For the other two cases with $\hat c_l=0.2$ and 0.5 in figure~\ref{result}(g,h), the corrected inlet concentration has limited influence on the model predictions, which is expected since the segregation flux is less than half of that for $\hat c_l=0.8$, and the particles entering the flowing layer are relatively well mixed. 

\begin{figure}
    \centerline{\includegraphics[width=5.0 in]{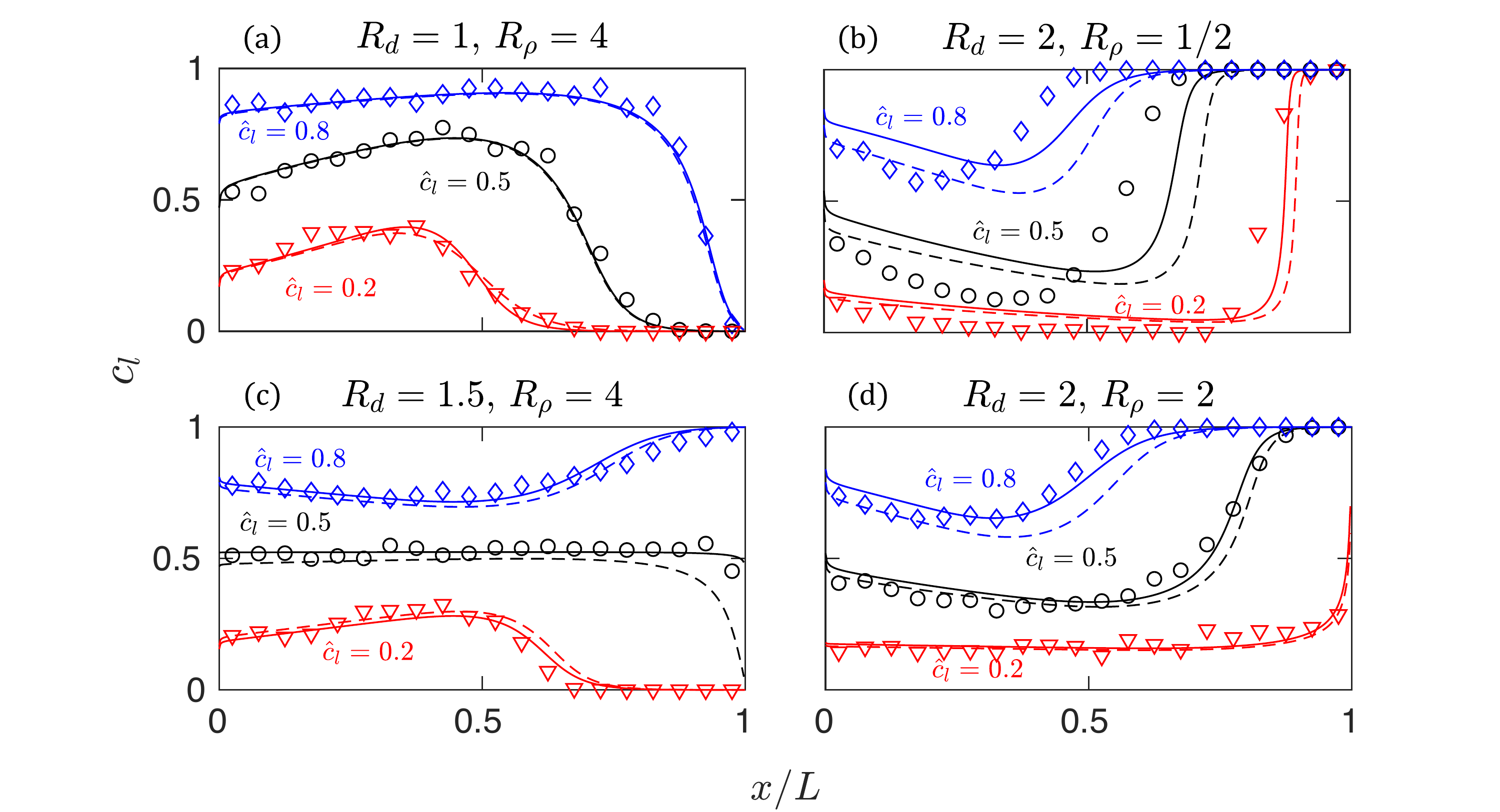}}
    \caption{
    Streamwise large particle concentration profiles at $z/\delta=-1$ for different feed concentrations with a variety of size and density ratios from simulation (symbols) and model (curves). Solid curves are model predictions with the inlet boundary condition $c_i(\tilde x=0)=\hat c_i$ while dashed curves are model predictions with a corrected inlet boundary condition measured from simulation, $c_{i} (\tilde x=0,\tilde z)=\langle c_{i}(\tilde x,\tilde z=-1)\rangle$.
     $q=15\,$cm$^2/$s, $\delta\approx1.5\,$cm, and $L=43\,$cm.
    }
    \label{result1}
\end{figure}

A more rigorous test of the continuum segregation model and the quadratic form of the segregation velocity of equation~(\ref{quadkappa}) is to consider flows with different combinations of size and density ratios ($R_d=1-2$ and $R_\rho=1/2-4$), a different flow geometry ($W=0.4\,$m), and a different feed rate ($q=15\,$cm$^2$/s) than those used to obtain the parameters of the segregation model, i.e. $A_i$ and $B_i$ in figure~\ref{coeff}.  
Similar to figure~\ref{result}(g-i), the predicted concentration profiles (curves) for particles deposited on the heap are compared with the simulation results (data points) in figure~\ref{result1}, but only for the large particle concentration so that three different feed concentrations can be more clearly shown for each case.
Among the four different cases, the $R_d=1$ and $R_\rho=4$ case reduces to density only segregation, for which the continuum segregation model works well, as shown in figure~\ref{result1}(a).
A more interesting case occurs when size and density differences enhance one another, as shown in figure~\ref{result1}(b) for $R_d=2$ and $R_\rho=1/2$.
Again, the model predictions match the simulation results reasonably well despite some discrepancies for $\hat c_l=0.5$ and 0.8 due to significant segregation in the feed-zone similar to that in figure~\ref{result}(i). 
Using the corrected inlet concentration determined from the simulations significantly reduces these discrepancies (solid curves).

For $R_d=1.5$ and $R_\rho=4$ in figure~\ref{result1}(c) and $R_d=2$ and $R_\rho=2$ in figure~\ref{result1}(d), the size and density segregation mechanisms oppose one another. 
In figure~\ref{result1}(c), the shape of the curves depends strongly on the inlet concentration, particularly at the downstream end of the heap (large $x/L$) where the large particle concentration goes to 1 for large $\hat c_l$ and 0 for small $\hat c_l$ in figure~\ref{result1}(c).
According to figure~\ref{phase}, the equilibrium concentration is $c_{l,eq}\approx 0.5$, and the results for $\hat c_l=0.5$ in figure~\ref{result1}(c) indeed show that the particles remain mixed throughout the entire length of the flowing layer.
Figure~\ref{result1}(d) corresponds to a case where $c_{l,eq}\approx 0.1$ [see figures~\ref{fluxit}(a) and \ref{phase}].
Again, particles remain mixed at $\hat c_l=0.2$ for almost the entire length of the flowing layer, as would be expected for this concentration that is very close to the equilibrium concentration.
On the other hand, for other values of $\hat c_l$, the particles segregate. 
Again the use of a corrected inlet concentration improves the match between the continuum model and the simulation results.
However, even without this correction, the agreement for all four cases in figure~\ref{result1} is remarkable given the wide range of $R_d$, $R_\rho$, and $\hat c_l$, not to mention the simplifying assumptions for the velocity profile and uniform flowing layer thickness. This shows that with appropriate model coefficients, the quadratic segregation velocity model of equation~(\ref{quadkappa}) can accurately predict bidisperse heap flow segregation not only for different feed concentrations, but also for different particle size and density ratios.

\section{Conclusions}
\label{section5}

When two segregation driving mechanisms (i.e., percolation and buoyancy) compete with each other, the segregation flux varies non-monotonically with particle concentration such that the segregation direction can reverse, depending on the relative concentration of the two species. 
This is different from size or density segregation alone, where the segregation velocity and flux never change sign \citep{schlick2015modeling,xiao2016modelling,jing2017micromechanical,jones2018asymmetric}.

Although the segregation behaviour of SD-systems is more complicated than S- or D-systems alone, it is possible to model the local concentration dependent rising and sinking behaviour. To do so, we extend a predictive model for either size or density segregation alone to bidisperse granular materials where particles can differ simultaneously in both size and density using a segregation velocity model that is quadratic in particle concentration (\ref{quadkappa}).
The two model coefficients can be determined from DEM simulations of quasi-2D bounded heap flow for a range of combinations of particle size and density ratios.
By incorporating the quadratic segregation velocity model into an advection-diffusion-segregation transport equation, the model predictions match well with simulation results.

The theoretical framework presented here is consistent with previous studies, which means the extended segregation velocity model of equation~(\ref{quadkappa}) should be applicable to not only quasi-2D bounded heap flow but also other 2D \citep{schlick2015granular,xiao2019continuum} and 3D \citep{isnergranular,isner2020axisymmetric} geometries with known flow kinematics.  
Moreover, continuous size distributions could be considered in this same framework for polydisperse segregation \citep{schlick2016continuum,deng2019modeling,gao2020broad}.
However, more research is needed to validate the model for a wider range of particle properties, particularly with respect to the effect of particle size and density ratios on the coefficients $A_i$ and $B_i$. 
Of particular value would be determining the functional dependence of $A_i$ and $B_i$ on the physical parameters of the mixture analogous to the dependence of $A$ in the linear segregation velocity model (\ref{wps}) on the logarithm of the particle size or volume ratio for S-systems \citep{schlick2015modeling,jones2020remarkable} and on the logarithm of the particle density ratio in D-systems \citep{xiao2016modelling}.
Some progress has been made in this direction for single intruder particles in SD-systems \citep{jing2020rising}, but connecting single intruder results to mixtures of particles like those considered here is not yet possible.
Nevertheless, the results in figures~\ref{result} and \ref{result1} indicate that the extended segregation velocity model can accurately predict combined size and density segregation. 

An interesting implication of the concentration dependent segregation direction is the stability of particle mixtures at the equilibrium concentration indicated in figure~\ref{phase}.
For heap flows like those in figure~\ref{flux}, reduced segregation flux near the equilibrium concentration increases the segregation timescale to be much greater than the timescale of heap flow kinematics. 
In such cases, particles deposit onto the heap without significant segregation occurring.
Or, in a more interesting situation, segregation occurs until the local concentration reaches the equilibrium concentration, $c_{l,eq}$, and then stops. 
Evidence of this appears in figure~\ref{flux}(a) where the data points reach a limit near the equilibrium concentration ($c_{l,eq}\approx 0.4$) when approached from either side of the equilibrium concentration. 
For example, blue data points for $\hat c_l=0.3$ fall in the range of $0\leq c_l \lessapprox 0.4$ because once the local concentration reaches $c_{l,eq}\approx 0.4$ the particles no longer segregate and remain mixed. 
As a result, only a small fraction of the data points are for a local concentration that exceeds $c_{l,eq}$.
A similar result occurs when $c_{l,eq}$ is approached from a higher concentration, {although it is not clearly evident in figure~\ref{flux}(a) because of how data points overlay one another.}
This could have interesting implications for other flow configurations such as rotating tumblers, where particle mixtures initially near the equilibrium concentration can evolve for sufficient time that an instability occurs in which a small perturbation of concentration results in segregation that pushes the local mixture concentration further away from the equilibrium concentration and further segregation continues.
This is beyond the scope of this study but deserves further investigation.

The problem of segregation in granular materials is too expansive to ever be declared fully solved for all possible mixtures and flows.  Advances in direct simulations will continue and then become routine, and new conceptual breakthroughs will occur that yield new insights into the physics of the problem. However, we seem to have reached a stable rung on the ladder of modelling segregation in terms of continuum models.  Models describing size or density driven segregation have been shown to capture--in three way comparisons--both experimental and computational results.  And now, in what may be regarded as nearly the last step in the evolution of such continuum models, we have shown that it is possible to simultaneously model size and density segregation.

\section*{Acknowledgements}

We thank Yi Fan, John Hecht, Anthony Thornton, Hongyi Xiao, and Lu Jing for valuable discussions. This material is based upon work supported by the National Science Foundation under Grant No. CBET-1929265.

\appendix

\renewcommand\thefigure{\thesection.\arabic{figure}}    
\setcounter{figure}{0}   

\renewcommand\thetable{\thesection.\arabic{table}}    
\setcounter{table}{0}   

\clearpage

\section{Experiment}\label{appendix}

\begin{figure}
    \centerline{\includegraphics[width=5.1 in]{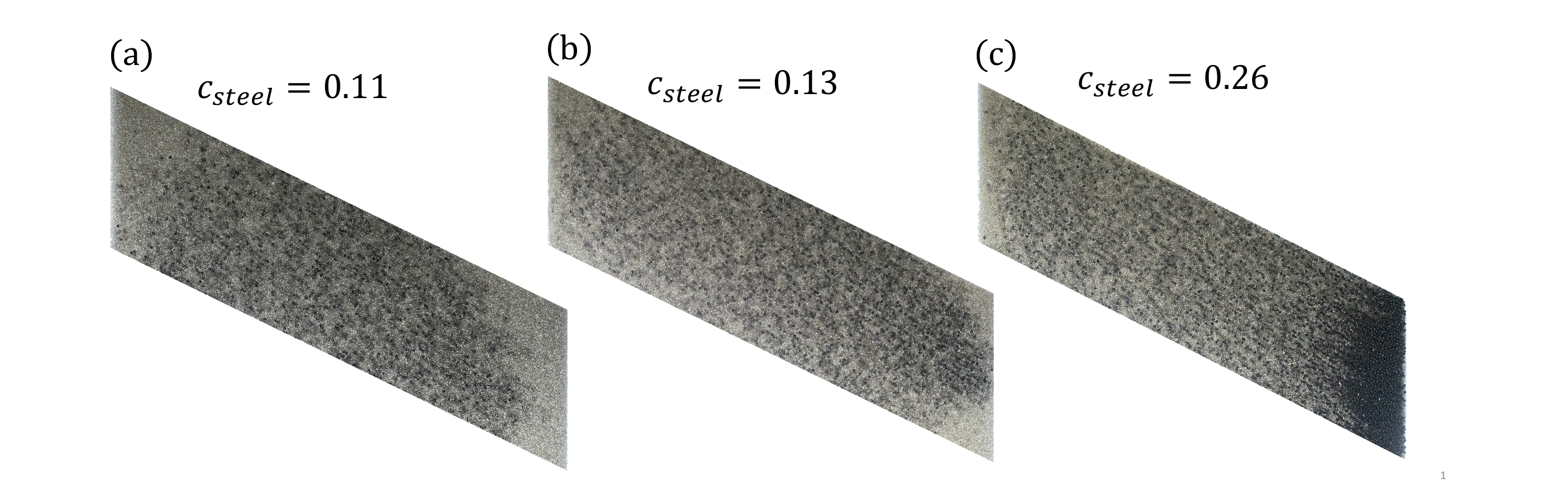}}
    \caption{Heap flow segregation for mixtures of large steel (dark) and small glass (light) particles with $R_d\approx 2$ and $R_\rho\approx 3$. Steel particles sink while glass particles rise for $c_{steel}=0.11$, as buoyancy overcomes percolation, leading to more glass particles depositing at the downstream end. In contrast, for $c_{steel}=0.26$ segregation reverses as percolation dominates over buoyancy. Particles remain relatively mixed for $c_{steel}=0.13$. 
     }
    \label{experiment}
\end{figure}

To validate the concentration dependent segregation direction observed in DEM simulations and predicted by the model, experiments were performed with large steel and small glass particles.  The experimental setup is similar to that in the simulation. The heap width between the bounding endwalls is $W=0.4$\,m. The gap between the two parallel glass plates is $T=1\,$cm. Particle mixtures are fed into the system by an auger feeder at a rate of $Q=12$\,cm$^3$/s. The width of the feed-zone is $W_f=4\,$cm. With the particle properties given in Table~\ref{table2}, the size ratio of large steel to small glass particles is $R_d\approx 2$ and the density ratio is $R_\rho\approx 3$.  Figure~\ref{experiment} shows images for different feed concentrations. The segregation direction reverses as the feed concentration of steel particles, $c_{steel}$, is increased from 0.11 to 0.26. 

\begin{table}
  \begin{center}
\def~{\hphantom{0}}
  \begin{tabular}{lccc}
      material  & ~~~color   &   ~~~~~diameter (mm) & ~~~density (g/cm$^3$) \\[3pt]
      glass   & ~~~light & ~~~~~1.45 $\pm$ 0.2 & ~~~2.58\\
      steel  & ~~~dark & ~~~~~2.98 $\pm$ 0.04 & ~~~7.84\\
  \end{tabular}
  \caption{Particle properties in experiments}
  \label{table2}
  \end{center}
\end{table}

\clearpage

\clearpage

\bibliographystyle{jfm}
\bibliography{jfm-instructions}

\end{document}